\documentclass[a4paper,12pt]{article}
\usepackage[colorlinks=true,linkcolor=blue,urlcolor=blue,filecolor=black,citecolor=blue,
pdfstartview=FitV,pdftitle={Phase transitions and quasinormal modes of Euler-Heisenberg-dilaton AdS black holes},
pdfsubject={Black-hole thermodynamics and quasinormal modes},pdfkeywords={AdS black holes, phase transitions, quasinormal modes},
hypertexnames=false,
pdfpagemode=UseNone,bookmarksopen=true]{hyperref}
\usepackage{graphicx}
\graphicspath{{figs_v9/}}
\usepackage{epstopdf}
\usepackage{amsmath,amsfonts,amssymb}
\usepackage{dcolumn}
\usepackage{slashed}
\usepackage{ulem}
\usepackage{float}
\usepackage{amsthm}
\usepackage{mathrsfs}
\usepackage{subfigure}
\usepackage{wrapfig}
\usepackage{indentfirst}
\usepackage{multirow}
\usepackage{appendix}
\usepackage{suffix}
\usepackage{color}
\WithSuffix\newcommand\footnote*[1]{\footnote[#1]{\addtocounter{footnote}{-1}}}

\makeatletter
\renewcommand\@makefnmark{\hbox{\@textsuperscript{\normalfont\@thefnmark}}}

\newcommand{\Rmnum}[1]{\expandafter\@slowromancap\romannumeral #1@}
\newcommand{\eq}[1]{Eq.(\ref{#1})}
\makeatother
\newcommand{\ud}{\,\mathrm{d}\,}

\begin{document}

\vspace{9mm}

\title{Phase transitions and quasinormal modes of Euler-Heisenberg-dilaton AdS black holes}

\begin{center}
{{{\Large {\bf Phase transitions and quasinormal modes of Euler-Heisenberg-dilaton AdS black holes }}}}\\[10mm]

{Yi-Qi Chen$^{1}$\footnote{chenyiqi0129@163.com}, De-Cheng Zou$^{1}$\footnote{Corresponding author: dczou@jxnu.edu.cn},
Ming Zhang$^{2}$\footnote{zhangming@xaau.edu.cn}, Xufen Zhang$^{3}$\footnote{xufenzhang0925@126.com}, \\ Bo Liu$^{4}$\footnote{fenxiao2001@163.com} and Rui-Hong Yue$^{5}$\footnote{rhyue@yzu.edu.cn}}\\[8mm]

{${}^1$School of Physics, Jiangxi Normal University, Nanchang 330022, China\\[0pt]}
{${}^2$Faculty of Science, Xihang University, Xi'an 710077 China\\[0pt]}
{${}^3$Key Laboratory of Low Dimensional Quantum Structures and Quantum Control of Ministry of Education,
Synergetic Innovation Center for Quantum Effects and Applications, and Department of Physics,
Hunan Normal University, Changsha, Hunan 410081, China\\[0pt]}
{${}^4$School of Arts and Sciences, Shaanxi University of Science and Technology, Xi'an, 710021, China\\[0pt]}
{${}^5$Center for Gravitation and Cosmology, College of Physical Science and Technology, Yangzhou University, Yangzhou 225009, China \\[0pt]}

\end{center}

\begin{abstract}
\indent
In this paper, we study phase transitions and scalar quasinormal modes(QNMs) of magnetically charged Euler--Heisenberg-dilaton AdS black holes in the extended phase space. We classify the horizon structures for all three signs of the EH coupling and derive the extended first law when the pressure and coupling are varied. For negative coupling, the physical branch exhibits a Van der Waals-like small/large black-hole transition. Using the improved asymptotic iteration method, we find two separated QNM trajectories below the critical pressure, whereas they merge into one continuous curve at the critical point. The GHS limit and the positive couplings considered here do not produce a finite-radius critical point.
\end{abstract}

\section{Introduction}
\label{intro}

Thermodynamic phase transitions of black holes in anti-de Sitter(AdS) spacetime have attracted much attention, mainly due to the AdS/CFT correspondence\cite{Maldacena:1997re,Gubser:1998bc,Witten:1998qj}. In the extended phase space, the cosmological constant is treated as thermodynamic pressure and the black-hole mass is understood as enthalpy\cite{Kastor:2009wy,Kubiznak:2012wp}, so charged AdS black holes can exhibit a liquid-gas-like small/large black-hole(SBH/LBH) phase transition. Extended-phase-space thermodynamics with nonlinear electromagnetic sources, higher-curvature gravity and reentrant transitions has been discussed for many black-hole models\cite{Gunasekaran:2012dq,Hendi:2012um,Zhao:2013oza,Zou:2013owa,Dehghani:2014caa,Hennigar:2015esa,Zhang:2014jfa,Altamirano:2014tva,Wei:2015iwa,Xu:2014kwa,Sadeghi:2016dvc,Hansen:2016ayo,Zeng:2016fsb,Hendi:2016usw,Xu:2015rfa,Zou:2016sab,Hendi:2017fxp,Rajagopal:2014ewa}. These results motivate a direct test of how a nonlinear matter model modifies the phase structure and whether the transition can be traced in dynamical perturbations.

The Euler--Heisenberg(EH) Lagrangian was introduced as an effective description of vacuum polarization in strong electromagnetic fields\cite{Heisenberg:1936nmg,Obukhov:2002xa}. Its gravitational coupling has produced electrically and magnetically charged black-hole solutions\cite{Yajima:2000kw,Ruffini:2013hia}. Rotating generalizations\cite{Breton:2019arv,Amaro:2022yew} and regular or multi-horizon geometries\cite{Guerrero:2020uhn,Nashed:2021ctg} have also been reported. An exact string-inspired solution in this class was recently obtained\cite{Bakopoulos:2024hah}. This solution carries a magnetic charge and contains two different metric functions, unlike the usual one-function Reissner--Nordstr\"om geometry. Its relation to dilaton black holes has been discussed in Refs.~\cite{Gao:2004gc,Gibbons:1987ps,Garfinkle:1990qj}, while perturbative properties were studied in Refs.~\cite{Ferrari:2000ep,Chen:2004zr,Shu:2004fj,Karimov:2018whx}. Particle motion and accretion in EH backgrounds have been investigated in Refs.~\cite{Yasir:2025npe,Jiang:2024njc,Vachher:2024ezs}; recent work also covers lensing, shadows and external-field perturbations\cite{Huang:2025jfa,Zhang:2025xqt}. In particular, the zero-EH-coupling limit is the magnetically charged Gibbons--Maeda--Garfinkle--Horowitz--Strominger(GHS) family, whose AdS dilaton continuation has been studied in Ref.~\cite{Gao:2004gc}. However, the influence of the effective EH-coupling sign on the horizon structure and Van der Waals-like behavior has not yet been analyzed systematically for its AdS extension.

Quasinormal modes(QNMs) are damped oscillations of perturbed black holes, and their frequencies and damping times encode the response of a black-hole geometry\cite{Konoplya:2011qq}. Their connection with thermodynamic phase transitions in AdS spacetime has been studied extensively\cite{Gubser:2000mm,Shen:2007xk,Mahapatra:2016dae,Rao:2007zzb,He:2010zb,Berti:2008xu,Koutsoumbas:2006xj,Zou:2014sja,Zou:2017juz,Zangeneh:2017rhc,Liu:2014gvf,Chabab:2016cem}. Recent studies have considered coupled gravito-electromagnetic perturbations, charged modified-gravity black holes, and AdS black holes through QNM spectra, greybody factors and thermodynamic observables\cite{Meng:2022oxg,Fu:2022cul,Lin:2024ubg}. Related Euler--Heisenberg backgrounds have been examined using axial, polar and de Sitter QNMs\cite{Zou:2025rbu,Li:2026gqi,Zhang:2026nog}. In particular, the slope of QNM frequencies changes differently on the two stable branches near a Van der Waals-like first-order transition, whereas this criterion becomes smooth at the second-order critical point. Motivated by these results, we ask whether this dynamical signature persists for the magnetically charged Euler--Heisenberg-dilaton AdS black hole with $A(r)\neq B(r)$.

In the present work, we classify the horizons for $\epsilon<0$, $\epsilon=0$ and $\epsilon>0$, and derive the thermodynamic quantities while varying both the cosmological pressure and the dimensionful EH coupling. We locate physical critical points by requiring two positive-temperature extrema to coalesce on the branch $K(r_+)>Q_{\mathrm{m}}^2$; the formal boundary $M^2\rightarrow0$ is not counted as a phase transition. For the QNM calculation, we use the improved asymptotic iteration method (AIM)\cite{Cho:2009cj,Cho:2011sf} together with ingoing-horizon and normalizable AdS boundary conditions.

The paper is organized as follows. Section~\ref{sol} presents the black-hole solution, the horizon structure and the extended thermodynamics, including the critical-point conditions for different $\epsilon$. Section~\ref{qnm} derives the scalar perturbation equation and the improved AIM, and follows the QNMs along first-order isobaric and isothermal paths. Section~\ref{critqnm} examines the critical isobar and isotherm, together with diagnostics for branch separation, coupling, mode dependence and critical scaling. Section~\ref{con} summarizes the results and states the limitation of the test-field analysis. Throughout this paper, we use the $(-,+,+,+)$ signature and set $G_N=\hbar=c=k_B=1$.

\section{Black hole solution, horizon structure and thermodynamics}
\label{sol}

\subsection{Black hole solution}
\label{solution}

We consider the Einstein-dilaton theory coupled to the nonlinear Euler--Heisenberg electromagnetic model, whose action is given by\cite{Bakopoulos:2024hah}
\begin{eqnarray}
&&{\cal S}=\frac{1}{16\pi}\int\ud^4x\sqrt{-g}\Big[R-2\nabla^\mu\phi\nabla_\mu\phi-\mathrm{e}^{-2\phi}F^2 \nonumber\\
&&\hspace{13mm}-f(\phi)\Big(2\alpha F^\mu_{~\nu}F^\nu_{~\rho}F^\rho_{~\delta}F^\delta_{~\mu}-\beta(F^2)^2\Big)-\mathfrak{V}(\phi)\Big].\label{action}
\end{eqnarray}
Here $F^2=F_{\mu\nu}F^{\mu\nu}$ and $\phi$ denotes the dilaton field. The constants $\alpha$ and $\beta$ control the nonlinear electromagnetic correction, and we choose the coupling function and dilaton potential as
\begin{eqnarray}
&&f(\phi)=-\Big[3\cosh(2\phi)+2\Big],\nonumber\\
&&\mathfrak{V}(\phi)=\frac{\Lambda}{3}\mathrm{e}^{-2\phi}+\frac{\Lambda}{3}\mathrm{e}^{2\phi}+\frac{4\Lambda}{3},\qquad \epsilon=\alpha-\beta,\label{coupling}
\end{eqnarray}
where $\Lambda<0$ is the cosmological constant. The difference $\epsilon=\alpha-\beta$ is the effective coupling that enters the black-hole solution.

For a static and spherically symmetric configuration, the metric, dilaton and magnetic gauge potential take the form\cite{Bakopoulos:2024hah}
\begin{eqnarray}
&&\mathrm{d}s^2=-A(r)\mathrm{d}t^2+\frac{\mathrm{d}r^2}{B(r)}+r^2\mathrm{d}\Omega_2^2,\label{metric}\\
&&A(r)=1-\frac{\sqrt{Q_{\mathrm{m}}^4+4M^2r^2}-Q_{\mathrm{m}}^2}{r^2}-\frac{2\epsilon Q_{\mathrm{m}}^4}{r^6}-\frac{\Lambda r^2}{3},\label{Afun}\\
&&B(r)=A(r)\frac{Q_{\mathrm{m}}^4+4M^2r^2}{4M^2r^2},\label{Bfun}\\
&&\phi(r)=-\frac{1}{2}\ln\Bigg[\frac{\sqrt{Q_{\mathrm{m}}^4+4M^2r^2}-Q_{\mathrm{m}}^2}{\sqrt{Q_{\mathrm{m}}^4+4M^2r^2}+Q_{\mathrm{m}}^2}\Bigg],\qquad \boldsymbol{A}_\mu=(0,0,0,Q_{\mathrm{m}}\cos\theta).\label{phisol}
\end{eqnarray}
Here $M$ and $Q_{\mathrm{m}}$ denote the mass and magnetic charge, respectively. The expression for $B(r)$ in \eq{Bfun} is algebraically equivalent to the form obtained from the field equations and makes the relation between $A(r)$ and $B(r)$ explicit. For $Q_{\mathrm{m}}\neq0$, one has $A(r)\neq B(r)$, so both functions enter the temperature and scalar wave equation.

When $\epsilon=0$ and $\Lambda=0$, the solution reduces to the magnetically charged Gibbons--Maeda--Garfinkle--Horowitz--Strominger (GHS) black hole introduced in Refs.~\cite{Gibbons:1987ps,Garfinkle:1990qj}. For $\Lambda<0$, it gives the corresponding AdS dilaton solution studied in Ref.~\cite{Gao:2004gc}. This limit is useful for comparison because, as shown below, its thermodynamics does not produce an RN--AdS-type Van der Waals critical point.

\subsection{Horizon structure and thermodynamic properties}
\label{thermo}

The event horizon $r_+$ is determined by $A(r_+)=0$. In the extended phase space, we identify the cosmological pressure as
\begin{eqnarray}
&&P=-\frac{\Lambda}{8\pi}.\label{pressure}
\end{eqnarray}
For later convenience, we define
\begin{eqnarray}
&&K(r_+)=r_+^2+Q_{\mathrm{m}}^2-\frac{2\epsilon Q_{\mathrm{m}}^4}{r_+^4}+\frac{8\pi P r_+^4}{3},\nonumber\\
&&K'(r_+)=2r_++\frac{8\epsilon Q_{\mathrm{m}}^4}{r_+^5}+\frac{32\pi P r_+^3}{3},\nonumber\\
&&\Delta(r_+)=\sqrt{K(r_+)^2-Q_{\mathrm{m}}^4}.\label{Kdef}
\end{eqnarray}
The horizon equation then gives
\begin{eqnarray}
&&K(r_+)=\sqrt{Q_{\mathrm{m}}^4+4M^2r_+^2}.\label{horizonk}
\end{eqnarray}
Hence, the mass is
\begin{eqnarray}
&&M=\frac{\Delta(r_+)}{2r_+}.\label{mass}
\end{eqnarray}
We select the positive-root branch, $K(r_+)>Q_{\mathrm{m}}^2$, in every calculation. The endpoint $M^2\rightarrow0$ produces formal temperature extrema, but it is not a physical turning point and is therefore excluded.

The sign of the short-distance term $-2\epsilon Q_{\mathrm{m}}^4/r^6$ controls the horizon structure. At large radius, all three cases approach
\begin{eqnarray}
&&A(r)=\frac{8\pi P}{3}r^2+1-\frac{2M}{r}+\frac{Q_{\mathrm{m}}^2}{r^2}+O(r^{-3}),\qquad r\rightarrow\infty,
\label{Ainfty}
\end{eqnarray}
Thus, $A(r)$ is positive in the asymptotic AdS region. At small radius, the leading behavior is
\begin{eqnarray}
&&A(r)\sim\begin{cases}
2|\epsilon|Q_{\mathrm{m}}^4/r^6, & \epsilon<0,\\
1-2M^2/Q_{\mathrm{m}}^2, & \epsilon=0,\\
-2\epsilon Q_{\mathrm{m}}^4/r^6, & \epsilon>0,
\end{cases}\qquad r\rightarrow0 .
\label{Aorigin}
\end{eqnarray}
For $\epsilon<0$, $A(r)$ tends to positive infinity at both endpoints. A sufficiently massive solution can therefore have two horizons separated by a region with $A(r)<0$, and their merger defines a finite-radius extremal black hole through
\begin{eqnarray}
&&A(r_{\rm ext})=0,\qquad A'(r_{\rm ext})=0 .\label{extremal}
\end{eqnarray}
For $Q_{\mathrm{m}}=0.4$, $P=0.01$ and $\epsilon=-0.5$, the corresponding numerical values are
\begin{eqnarray}
&&r_{\rm ext}=0.724148,\qquad M_{\rm ext}=0.541582,\qquad M=0.7:\nonumber\\
&&\hspace{10mm}(r_-,r_+)=(0.534873,1.13068),\label{negativehorizon}
\end{eqnarray}
whereas $M=0.45$ gives no horizon. The finite extremal mass therefore separates the horizonless and two-horizon solutions.

For $\epsilon=0$, \eq{Aorigin} gives the limiting value
\begin{eqnarray}
&&A(0)=1-\frac{2M^2}{Q_{\mathrm{m}}^2},\qquad M_0=\frac{Q_{\mathrm{m}}}{\sqrt{2}}=0.282843 .\label{ghsthreshold}
\end{eqnarray}
The threshold occurs at the zero-radius boundary rather than at a finite extremal radius. For $M=0.45$, the outer horizon is $r_+=0.666621$. When $\epsilon>0$, $A(r)$ tends to negative infinity at the origin and to positive infinity at infinity, so a positive-mass solution has one outer horizon. The sign of $\epsilon$ must therefore be retained when selecting thermodynamic branches.

\begin{figure}[htb]
\centering
\subfigure[a]{
\includegraphics[width=1.85in]{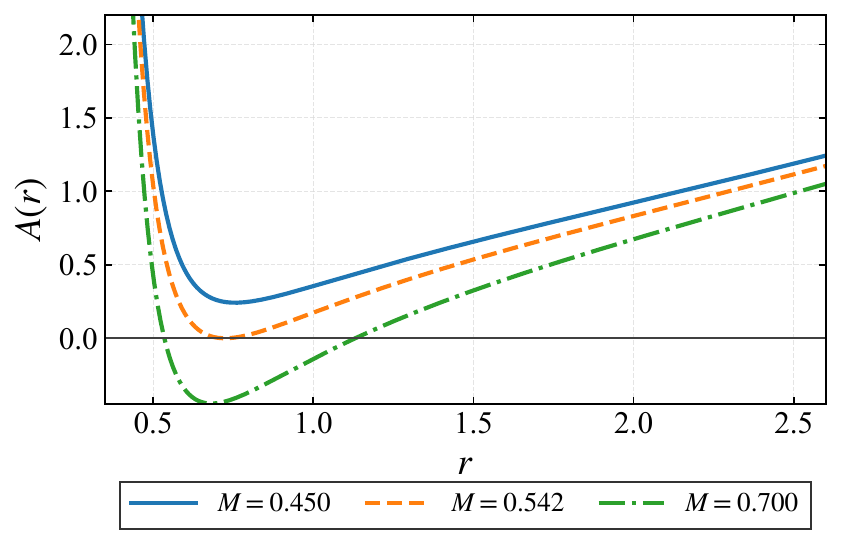}}
\hfill
\subfigure[b]{
\includegraphics[width=1.85in]{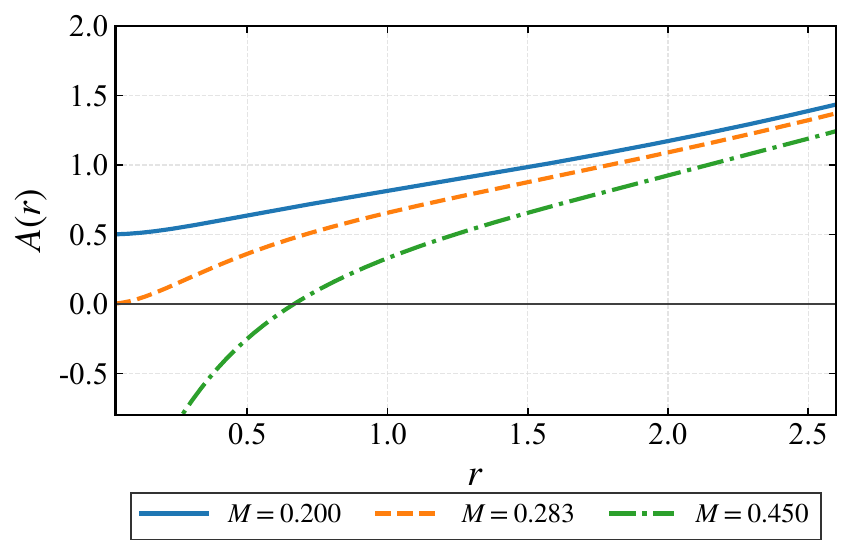}}
\hfill
\subfigure[c]{
\includegraphics[width=1.85in]{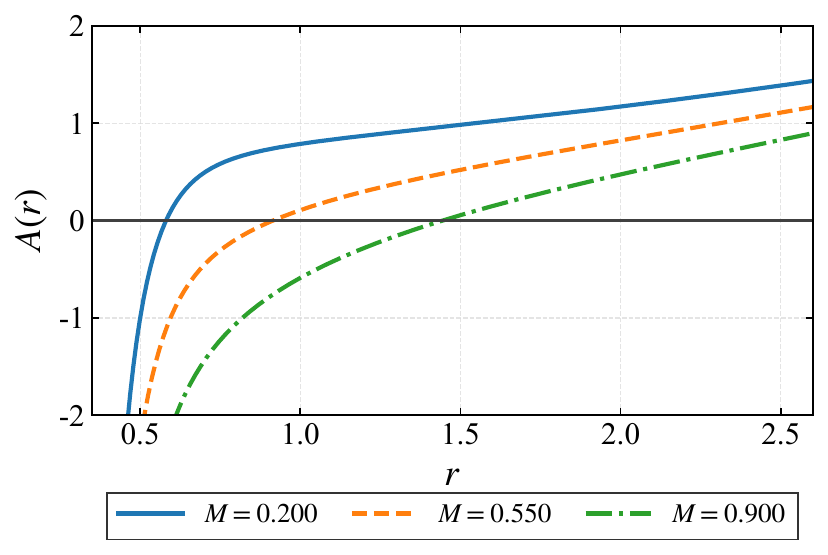}}
\caption{The metric function $A(r)$ as a function of $r$, with fixed $Q_{\mathrm{m}}=0.4$ and $P=0.01$. Panels (a), (b), and (c) correspond to $\epsilon=-0.5$, $0$ and $0.5$, with the mass values specified in the text, respectively. The sign of $\epsilon$ determines whether the geometry is horizonless, extremal or single-horizon in the displayed examples.}\label{fighorizon}
\end{figure}

The profiles are shown in Fig.~\ref{fighorizon}. In panel (a), $M=0.45$, $0.541582$ and $0.7$ give horizonless, degenerate and two-horizon solutions, respectively. Panel (b) displays the zero-coupling threshold and the outer horizon, while panel (c) displays the single horizon for positive coupling. These profiles determine the branches used below.

For the two-function metric, the Hawking temperature can be written as
\begin{eqnarray}
&&T=\frac{\sqrt{A'(r_+)B'(r_+)}}{4\pi}
 =\left(\frac{\partial M}{\partial S}\right)_{P,Q_{\mathrm{m}},\epsilon}
 =\frac{r_+K(r_+)K'(r_+)-\Delta(r_+)^2}{4\pi r_+^3\Delta(r_+)}.\label{temperature}
\end{eqnarray}
The derivatives in this expression are taken at fixed $P$, $Q_{\mathrm{m}}$ and $\epsilon$. The entropy obeys the area law. Differentiating \eq{mass} gives the thermodynamic volume, the magnetic potential and the quantity conjugate to $\epsilon$,
\begin{eqnarray}
&&S=\pi r_+^2,\qquad V=\frac{4\pi K(r_+)r_+^3}{3\Delta(r_+)},\label{SV}\\
&&\Phi_{\mathrm{m}}=\frac{K(r_+)\left(2Q_{\mathrm{m}}-8\epsilon Q_{\mathrm{m}}^3/r_+^4\right)-2Q_{\mathrm{m}}^3}{2r_+\Delta(r_+)},\nonumber\\
&&\Psi_{\epsilon}=-\frac{K(r_+)Q_{\mathrm{m}}^4}{r_+^5\Delta(r_+)}.\label{potentials}
\end{eqnarray}
Here $S$, $V$, $\Phi_{\mathrm{m}}$ and $\Psi_{\epsilon}$ denote the entropy, volume, magnetic potential and potential conjugate to $\epsilon$, respectively. We regard $M$ as the enthalpy and vary $\epsilon$ together with $P$ and $Q_{\mathrm{m}}$. The extended first law is
\begin{eqnarray}
&&\mathrm{d}M=T\mathrm{d}S+\Phi_{\mathrm{m}}\mathrm{d}Q_{\mathrm{m}}+V\mathrm{d}P+\Psi_{\epsilon}\mathrm{d}\epsilon,\label{firstlaw}
\end{eqnarray}
The independent variations are $S$, $Q_{\mathrm{m}}$, $P$ and $\epsilon$. Using $[S]=[\epsilon]=L^2$, $[Q_{\mathrm{m}}]=L$ and $[P]=L^{-2}$, we obtain the Smarr relation
\begin{eqnarray}
&&M=2TS+\Phi_{\mathrm{m}}Q_{\mathrm{m}}-2PV+2\Psi_{\epsilon}\epsilon.\label{smarr}
\end{eqnarray}

In the canonical ensemble, the Gibbs free energy is
\begin{eqnarray}
&&G=M-TS=\frac{3\Delta(r_+)^2-r_+K(r_+)K'(r_+)}{4r_+\Delta(r_+)}\nonumber\\
&&\hspace{8mm}=\frac{3\left[K(r_+)^2-Q_{\mathrm{m}}^4\right]-r_+K(r_+)K'(r_+)}{4r_+\sqrt{K(r_+)^2-Q_{\mathrm{m}}^4}}.\label{gibbs}
\end{eqnarray}
The heat capacity at fixed pressure, charge and coupling is
\begin{eqnarray}
&&C_P=T\left(\frac{\partial S}{\partial T}\right)_{P,Q_{\mathrm{m}},\epsilon}
 =\frac{2\pi r_+T}{(\partial T/\partial r_+)_P}.\label{cp}
\end{eqnarray}
The sign of $C_P$ distinguishes locally stable and unstable parts of an isobar. At fixed $Q_{\mathrm{m}}$ and $\epsilon$, \eq{gibbs} gives
\begin{eqnarray}
&&\mathrm{d}G=-S\mathrm{d}T+V\mathrm{d}P,
\label{gibbsdifferential}
\end{eqnarray}
An equal-$G$ crossing of two branches at fixed temperature locates a first-order transition, whereas a divergence of $C_P$ marks an extremum of $T(r_+)$ at fixed $P$. Here $T(P,r_+)$ is implicit through $K$ and $\Delta$, unlike the usual Reissner--Nordstr\"om--AdS expression. We solve \eq{temperature} numerically for $P(r_+,T)$ on the physical branch and use $G$ to distinguish a swallowtail from a single temperature minimum.

\subsection{Phase behavior for different EH couplings}
\label{pv}

For a physical Van der Waals-like critical point, two positive-temperature extrema of the isobaric $T-r_+$ curve must coalesce at fixed pressure. We impose
\begin{eqnarray}
&&\left(\frac{\partial T}{\partial r_+}\right)_{P}=0,\qquad \left(\frac{\partial^2 T}{\partial r_+^2}\right)_{P}=0.\label{critical}
\end{eqnarray}
The derivatives are evaluated on the positive-root branch $K(r_+)>Q_{\mathrm{m}}^2$, where $M$ is real and positive. This condition is imposed in every scan, so the formal zero-mass endpoint is not counted as a temperature extremum. 

We parameterize the critical quantities by
\begin{eqnarray}
&&\kappa_c=K(r_c),\qquad \Delta_c=\sqrt{\kappa_c^2-Q_{\mathrm{m}}^4}.\label{criticaldefs}
\end{eqnarray}
The relation $K(r_c)=\kappa_c$ then gives
\begin{eqnarray}
&&P_c=\frac{3}{8\pi r_c^4}\left(\kappa_c-r_c^2-Q_{\mathrm{m}}^2+\frac{2\epsilon Q_{\mathrm{m}}^4}{r_c^4}\right),\nonumber\\
&&T_c=\frac{3\kappa_c^2-2\kappa_c r_c^2-4\kappa_c Q_{\mathrm{m}}^2
+16\epsilon\kappa_cQ_{\mathrm{m}}^4/r_c^4+Q_{\mathrm{m}}^4}
 {4\pi r_c^3\Delta_c}.\label{criticalvalues}
\end{eqnarray}
The radius is fixed by the double-root conditions, and Eq.~\eqref{criticalvalues} gives the pressure and temperature of the critical point, $P_c$ and $T_c$. We retain a root only when $r_c>0$, $P_c>0$, $T_c>0$ and $\Delta_c>0$. The zero-coupling case can be settled analytically before the numerical search. Set
\begin{eqnarray}
&&x=r_+^2,\qquad a=\frac{8\pi P}{3}.\label{ghsvars}
\end{eqnarray}

For $\epsilon=0$, Eq.~\eqref{temperature} reduces to
\begin{eqnarray}
&&T_0(x)=\frac{1+2aQ_{\mathrm{m}}^2+4ax+3a^2x^2}
 {4\pi\sqrt{(1+ax)(x+2Q_{\mathrm{m}}^2+ax^2)}},\nonumber\\
&&\frac{\partial T_0}{\partial x}=\frac{E(x)}{8\pi H(x)^{3/2}},\qquad
H(x)=(1+ax)(x+2Q_{\mathrm{m}}^2+ax^2),\label{ghsstat}
\end{eqnarray}
where
\begin{eqnarray}
&&E(x)=3a^4x^4+8a^3x^3+6a^2(1+2aQ_{\mathrm{m}}^2)x^2+24a^2Q_{\mathrm{m}}^2x
 -1+12aQ_{\mathrm{m}}^2-4a^2Q_{\mathrm{m}}^4,\nonumber\\
&&E'(x)=12a^4x^3+24a^3x^2+12a^2(1+2aQ_{\mathrm{m}}^2)x+24a^2Q_{\mathrm{m}}^2>0.\label{ghsmono}
\end{eqnarray}
For $x>0$ and $P>0$, $E(x)$ is strictly increasing and has at most one positive zero. At such a zero,
\begin{eqnarray}
&&\frac{\partial^2T_0}{\partial x^2}=\frac{E'}{8\pi H^{3/2}}>0,\qquad
\frac{\partial^2T_0}{\partial r_+^2}=4x\frac{\partial^2T_0}{\partial x^2}>0.\label{ghssecond}
\end{eqnarray}
The two conditions in Eq.~\eqref{critical} therefore cannot hold at finite $r_+$, and the GHS limit has no Van der Waals critical point.

For nonzero $\epsilon$, we determine the roots from the functions $K(r_+)$ and $\Delta(r_+)$ by defining
\begin{eqnarray}
&&{\cal N}=r_+K K'-\Delta^2,\qquad
{\cal F}=r_+\Delta {\cal N}'-{\cal N}\left(3\Delta+r_+\Delta'\right),\nonumber\\
&&T=\frac{{\cal N}}{4\pi r_+^3\Delta},\qquad
\left(\frac{\partial T}{\partial r_+}\right)_P
 =\frac{{\cal F}}{4\pi r_+^4\Delta^2}.\label{criticalpolynomials}
\end{eqnarray}
Here and below, the prime denotes a derivative at fixed $P$, $Q_{\mathrm{m}}$ and $\epsilon$. At a stationary point, direct differentiation gives
\begin{eqnarray}
&&\left.\left(\frac{\partial^2T}{\partial r_+^2}\right)_P\right|_{{\cal F}=0}
 =\frac{{\cal F}'}{4\pi r_+^4\Delta^2}.\label{criticalderivatives}
\end{eqnarray}
The critical radius is the common positive root of ${\cal F}=0$ and ${\cal F}'=0$ on the physical branch. We then evaluate $P_c$ and $T_c$ from Eq.~\eqref{criticalvalues}. After eliminating $\kappa_c$, one algebraic equation for $r_c$ remains, and roots with $\Delta=0$ or at the radial cutoff are discarded.

Near a candidate point, set $\delta r=r_+-r_c$ and $\delta P=P-P_c$. The temperature expansion is
\begin{eqnarray}
&&\delta T=T_{P,c}\delta P+T_{rP,c}\delta r\,\delta P
 +\frac{1}{6}T_{rrr,c}\delta r^3
 +O(\delta P^2,\delta r^2\delta P,\delta r^4),\nonumber\\
&&\delta r_{\pm}=\pm\left(-\frac{2T_{rP,c}}{T_{rrr,c}}\delta P\right)^{1/2}+O(\delta P),
\qquad (\delta P<0),\label{criticalexpansion}
\end{eqnarray}
where subscripts denote derivatives at the candidate point. Below $P_c$, two extrema occur when the square-root argument is positive. At $P=P_c$, they merge and the isobar develops a cubic horizontal inflection. This criterion does not require an explicit elementary equation of state.

The sign of $\epsilon$ fixes the short-distance behavior in the remaining cases. For $\epsilon>0$,
\begin{eqnarray}
&&K(r_+)\sim-\frac{2\epsilon Q_{\mathrm{m}}^4}{r_+^4},\qquad r_+\rightarrow0.\label{Kpositiveorigin}
\end{eqnarray}
The physical branch then starts at a finite radius where $K=Q_{\mathrm{m}}^2$, and its positive-temperature part returns to the large-radius behavior
\begin{eqnarray}
&&T=2Pr_++O(r_+^{-1}),\qquad r_+\rightarrow\infty .\label{largeradiusT}
\end{eqnarray}
The function ${\cal F}$ has one positive-temperature zero for each tested positive coupling $\epsilon=0.01,0.1,0.5,1,2$, and ${\cal F}'>0$ there, so each zero is a single minimum. 

\begin{figure}[htb]
\centering
\begin{minipage}[t]{1.95in}\centering
\includegraphics[width=1.95in]{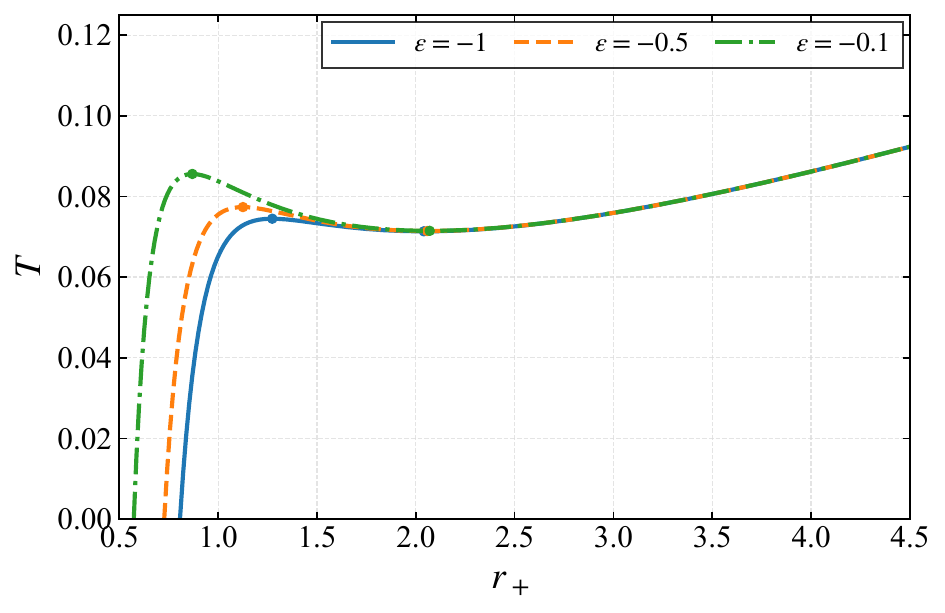}\\[-1mm]
{\small (a) $\epsilon<0$}
\end{minipage}\hfill
\begin{minipage}[t]{1.95in}\centering
\includegraphics[width=1.95in]{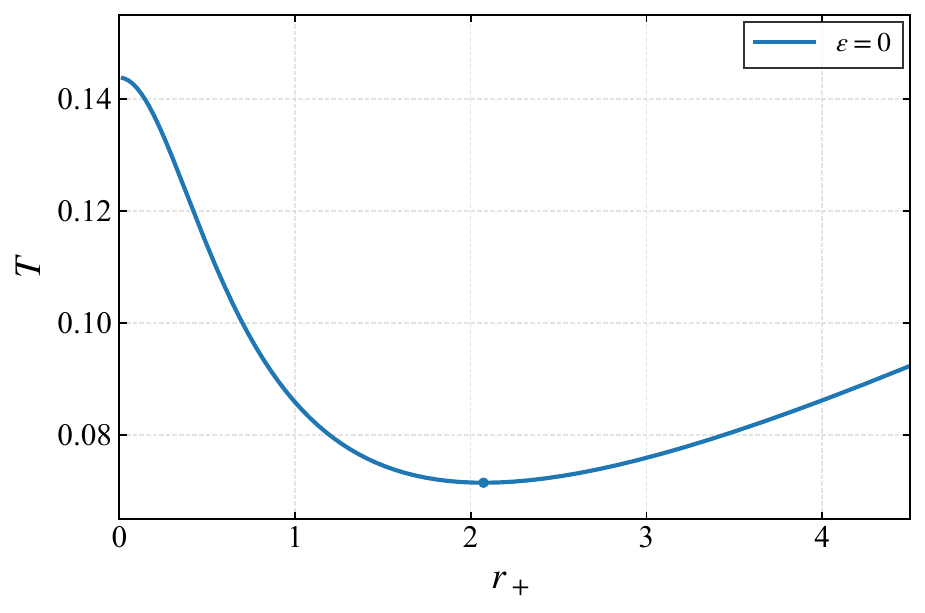}\\[-1mm]
{\small (b) $\epsilon=0$}
\end{minipage}\hfill
\begin{minipage}[t]{1.95in}\centering
\includegraphics[width=1.95in]{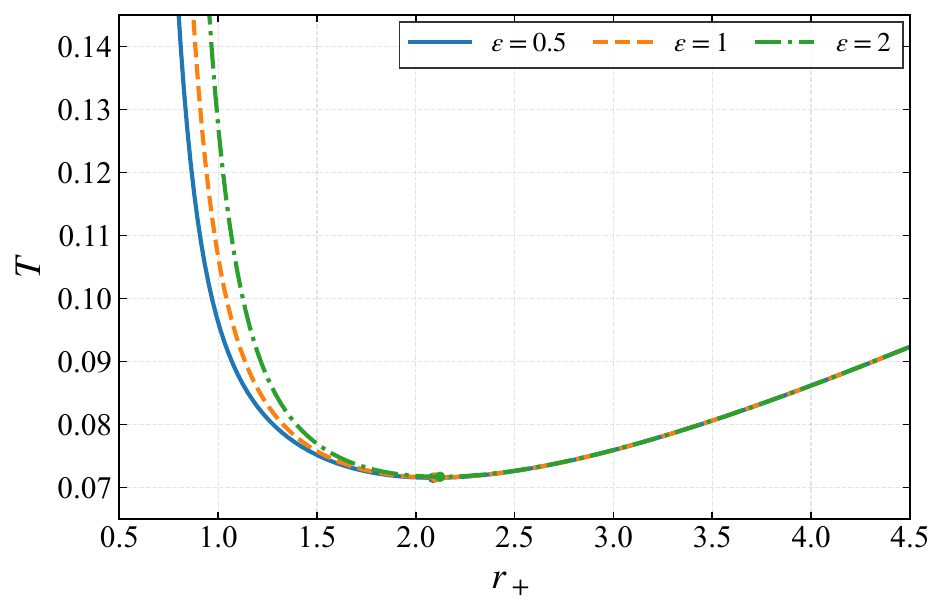}\\[-1mm]
{\small (c) $\epsilon>0$}
\end{minipage}
\caption{The Hawking temperature $T$ as a function of $r_+$, with fixed $Q_{\mathrm{m}}=0.4$ and numerical pressure $P=0.00830944$. Panels (a), (b), and (c) correspond to $\epsilon<0$, $\epsilon=0$, and $\epsilon>0$, respectively; the numerical values of the curves are written inside the corresponding plots. For the reference case $\epsilon=-0.5$, this pressure equals $0.6P_c$, whereas the other panels have different ratios $P/P_c$. The negative-coupling curves retain a maximum--minimum pair, while the zero and positive cases show only one minimum.}\label{figtempsigns}
\end{figure}

\begin{figure}[htb]
\centering
\begin{minipage}[t]{1.95in}\centering
\includegraphics[width=1.95in]{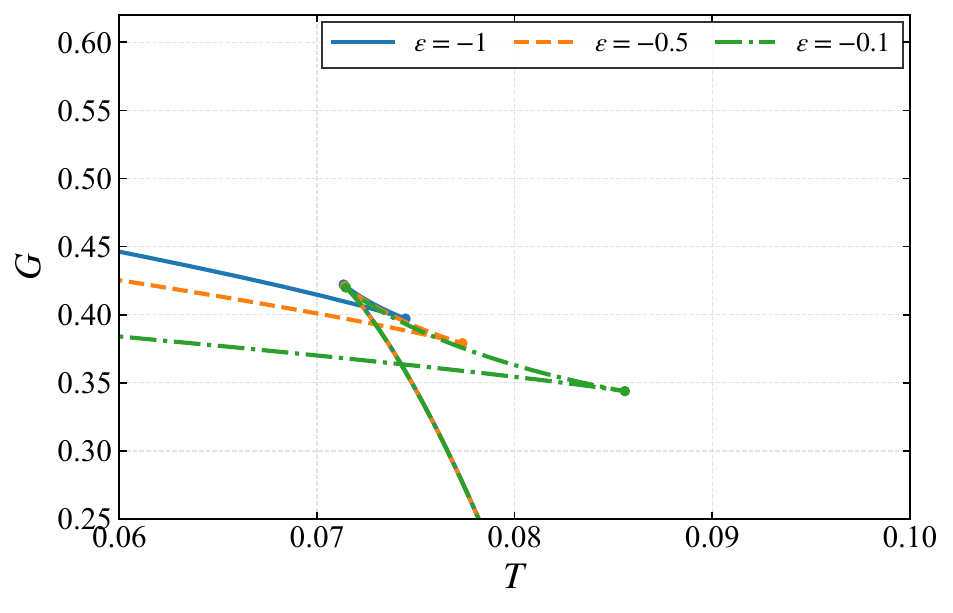}\\[-1mm]
{\small (a) $\epsilon<0$}
\end{minipage}\hfill
\begin{minipage}[t]{1.95in}\centering
\includegraphics[width=1.95in]{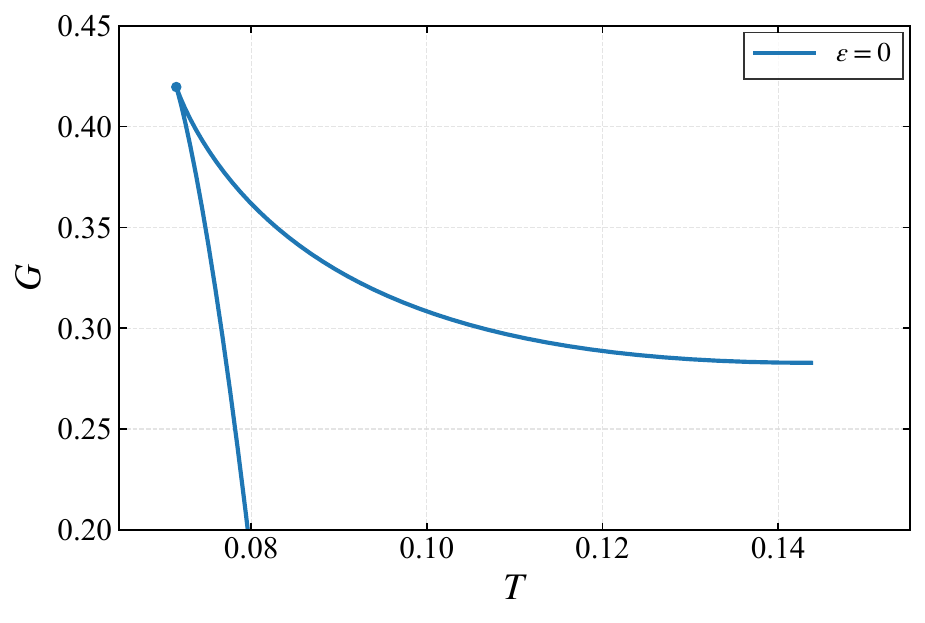}\\[-1mm]
{\small (b) $\epsilon=0$}
\end{minipage}\hfill
\begin{minipage}[t]{1.95in}\centering
\includegraphics[width=1.95in]{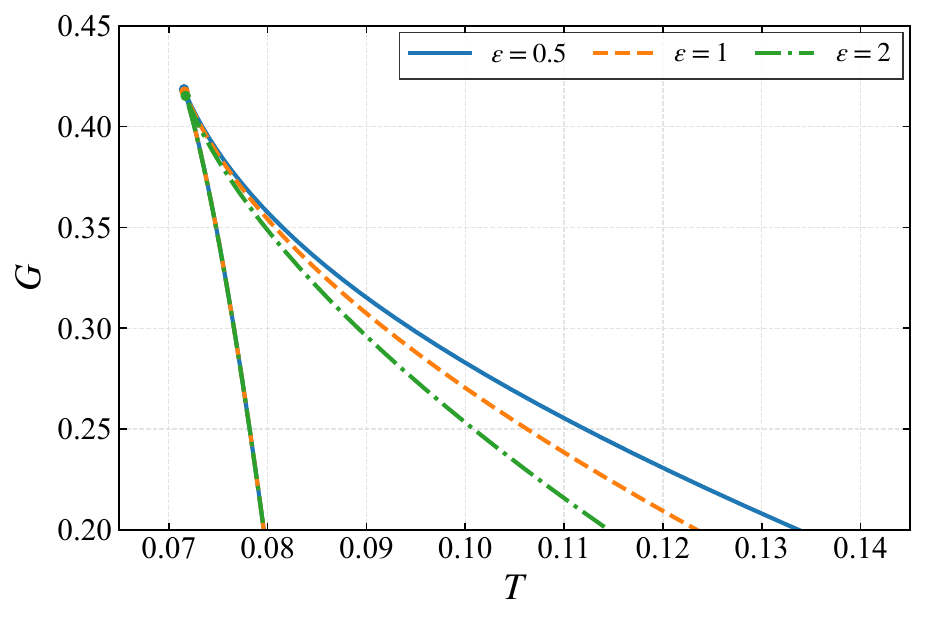}\\[-1mm]
{\small (c) $\epsilon>0$}
\end{minipage}
\caption{The Gibbs free energy $G$ as a function of $T$, with fixed $Q_{\mathrm{m}}=0.4$ and $P=0.00830944$. Panels (a), (b), and (c) correspond to $\epsilon<0$, $\epsilon=0$, and $\epsilon>0$, respectively; the numerical values of the curves are written inside the corresponding plots. Only the negative-coupling case develops a swallowtail and an equal-$T$, equal-$G$ crossing.}\label{figgibbssigns}
\end{figure}

For $\epsilon<0$, we set $e=|\epsilon|$. Then
\begin{eqnarray}
&&K(r_+)\sim\frac{2eQ_{\mathrm{m}}^4}{r_+^4},\qquad
T=-\frac{5eQ_{\mathrm{m}}^4}{2\pi r_+^7}+O(r_+^{-3}),\qquad r_+\rightarrow0 .\label{smallradiusT}
\end{eqnarray}
The temperature first crosses zero and can then develop a maximum and a minimum before approaching $2Pr_+$. For $Q_{\mathrm{m}}=0.4$, the equations ${\cal F}=0={\cal F}'$ give one admissible root for $\epsilon=-0.5$,
\begin{eqnarray}
&&r_c=1.34693,\qquad P_c=0.0138491,\qquad T_c=0.0904713.\label{criticalnumeric}
\end{eqnarray}
At this point,
\begin{eqnarray}
&&\Delta_c^2=K(r_c)^2-Q_{\mathrm{m}}^4>0. \label{criticalpositivity}
\end{eqnarray}
The residuals of Eq.~\eqref{criticalderivatives} are below $2\times10^{-8}$. At $P=0.6P_c$, the two temperature extrema are listed in Eq.~\eqref{turningvalues}, whereas the positive-coupling cases give only one minimum. Hence, the negative EH coupling supplies the two branches required for a Van der Waals-like transition, but the positive couplings tested here do not produce a finite-radius critical point.
For comparison, the other negative couplings give
\begin{eqnarray}
&&\epsilon=-1:~~(r_c,P_c,T_c)=(1.48938,0.0116781,0.0834539),\nonumber\\
&&\epsilon=-0.1:~~(r_c,P_c,T_c)=(1.07728,0.0197503,0.106653).\label{negativecriticalvalues}
\end{eqnarray}
An admissible critical root persists over the negative-coupling values used in Figs.~\ref{figtempsigns} and \ref{figgibbssigns}; as $\epsilon$ approaches zero from below, $r_c$ decreases while $P_c$ and $T_c$ increase.

We solve the root conditions simultaneously after imposing $K(r_c)>Q_{\mathrm{m}}^2$,
\begin{eqnarray}
&&{\cal F}(r_c,P_c)=0,\qquad \partial_{r_+}{\cal F}(r_c,P_c)=0,\label{numericrootconditions}
\end{eqnarray}
and then evaluate Eq.~\eqref{criticalvalues}. The residuals are evaluated from the original temperature derivatives. Inverting the temperature relation to $P=P(r_+,T)$ gives the same critical point. Implicit differentiation gives
\begin{eqnarray}
&&\left(\frac{\partial P}{\partial r_+}\right)_T=-\frac{(\partial T/\partial r_+)_P}{(\partial T/\partial P)_{r_+}},\qquad
\left(\frac{\partial^2P}{\partial r_+^2}\right)_T=-\frac{(\partial^2T/\partial r_+^2)_P}{(\partial T/\partial P)_{r_+}},\label{implicitcritical}
\end{eqnarray}
at a stationary point. The two descriptions are equivalent when $(\partial T/\partial P)_{r_+}\neq0$, so no separate isothermal criticality condition is needed.

We next compare the three signs of $\epsilon$. The numerical scan uses
\begin{eqnarray}
&&Q_{\mathrm{m}}=0.4,\qquad 0.02\leq r_+\leq50,\nonumber\\
&&10^{-4}\leq P\leq1,\qquad T>0.\label{scanrange}
\end{eqnarray}
where the last condition is imposed before evaluating the derivatives. For $\epsilon=-0.5$, this procedure gives the physical critical point quoted in Eq.~\eqref{criticalnumeric}. At the representative subcritical pressure $P=0.6P_c=0.00830944$, the two temperature extrema for $\epsilon=-0.5$ are located at
\begin{eqnarray}
&& (r_+,T)=(1.12507,0.0773561),\qquad (2.05755,0.0714132).\label{turningvalues}
\end{eqnarray}

By contrast, the same pressure gives one physical extremum for $\epsilon=0$ and $\epsilon=0.5$. A scan of \eq{critical} over \eq{scanrange} and 60 logarithmically spaced pressures gives at most one positive-temperature stationary point in either case. The GHS limit has one stationary point at 42 pressures and none at the other 18, whereas the $\epsilon=0.5$ case has one stationary point at all 60 pressures. Scans at $\epsilon=0.01,0.1,1$ and $2$ give the same pattern. Within the tested ranges of $Q_{\mathrm{m}}$, $r_+$, $P$ and $\epsilon$, neither $\epsilon=0$ nor $\epsilon>0$ develops the max--min pair required for a Van der Waals-like critical point.

Formal stationary solutions approaching $r_+\rightarrow0$ for $\epsilon=0$ are excluded because they do not form two extrema and move with the radial cutoff. Figure~\ref{figtempsigns} tests this result away from the representative values. At this pressure, all three negative-coupling curves retain two extrema, whereas the positive-coupling curves retain one minimum. Panel (a) contains $\epsilon=-1,-0.5,-0.1$, panel (b) contains $\epsilon=0$, and panel (c) contains $\epsilon=0.5,1,2$. Because $P_c$ varies with $\epsilon$, the fixed pressure corresponds to $P/P_c=0.7115$, $0.6000$ and $0.4207$ for $\epsilon=-1$, $-0.5$ and $-0.1$, respectively.

The temperature curves in Fig.~\ref{figtempsigns} also determine the local stability. Since $C_P=T(\partial S/\partial T)_P$, its sign follows $(\partial T/\partial r_+)_P$. For negative coupling, two extrema divide the branch into stable small, unstable intermediate and stable large black holes. For zero and positive coupling, one minimum separates an unstable small-radius part from a stable large-radius part, which is insufficient for SBH/LBH coexistence. The global test is shown in Fig.~\ref{figgibbssigns}, with the same coupling values as Fig.~\ref{figtempsigns}; its panel labels and in-panel line-style legends follow the same convention.

\begin{figure}[htb]
\centering
\makebox[\linewidth][c]{%
\subfigure[a]{
\includegraphics[width=2.8in]{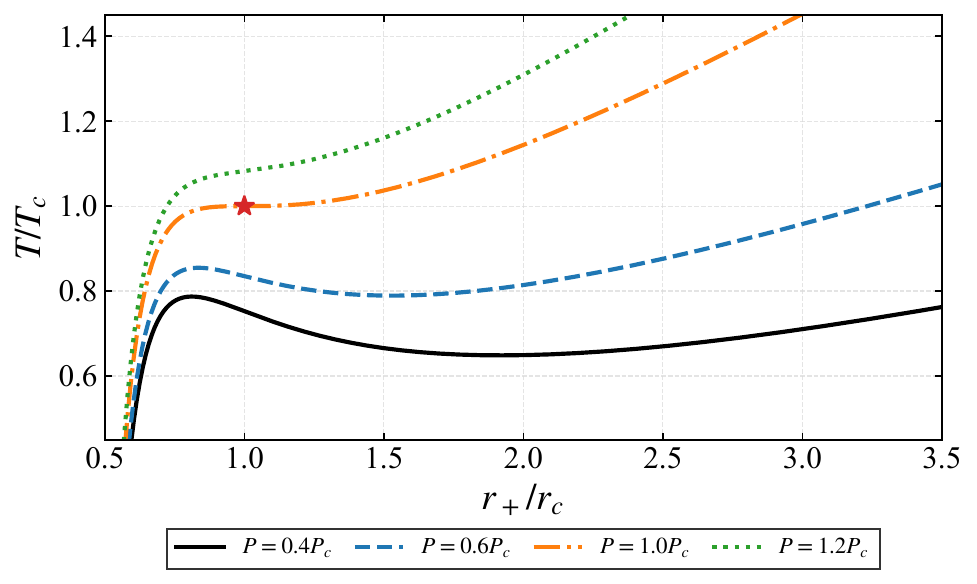}}
\hspace{0.10in}%
\subfigure[b]{
\includegraphics[width=2.8in]{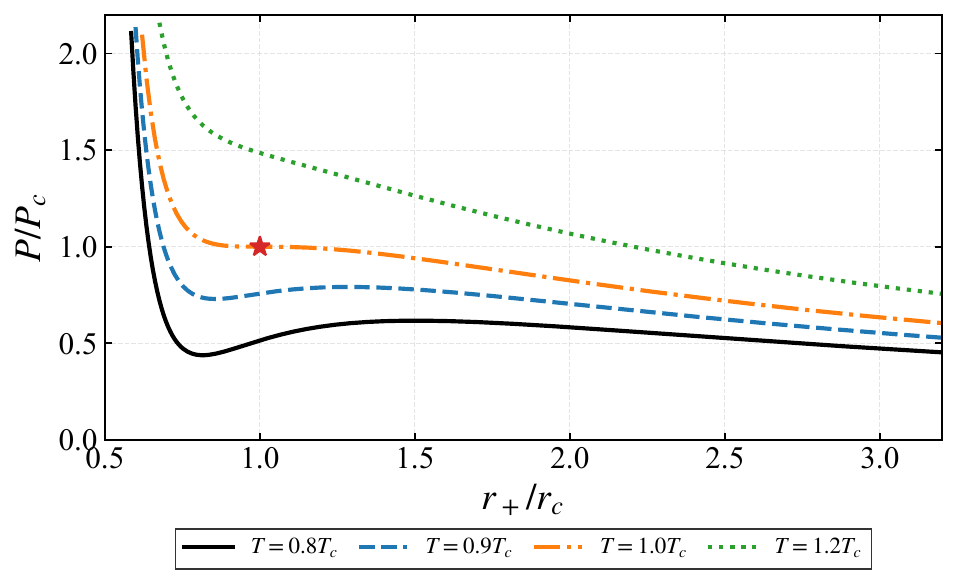}}}%
\caption{The temperature (a) and pressure (b) as functions of $r_+/r_c$, with fixed $\epsilon=-0.5$ and $Q_{\mathrm{m}}=0.4$. The curves in panels (a) and (b) correspond to $P/P_c=0.4,0.6,1.0,1.2$ and $T/T_c=0.8,0.9,1.0,1.2$, respectively. The oscillatory branches occur below the critical values, whereas the supercritical curves are monotonic.}\label{figthermopaths}
\end{figure}

The Gibbs curves in Fig.~\ref{figgibbssigns} provide the corresponding global-stability test. For $\epsilon=-0.5$, the swallowtail contains two states with the same temperature and Gibbs free energy. The equal-$T$, equal-$G$ construction gives
\begin{eqnarray}
&&T_*=0.0729629,\qquad r_s=0.951028,\qquad r_l=2.575539,\nonumber\\
&&G_*=0.392840,\label{coexistence}
\end{eqnarray}
where $r_s$ and $r_l$ are the coexisting small and large horizon radii. The Gibbs free energy is continuous at $T_*$, whereas $\partial G/\partial T=-S$ changes discontinuously, identifying a first-order transition below the endpoint \eq{criticalnumeric}. For $\epsilon=0$ and for the positive couplings tested here, no swallowtail appears and no two radii satisfy the equal-$T$, equal-$G$ conditions, so the transition is absent in these cases.
We also perform the two thermodynamic scans separately. In the isobaric process, $P$ is fixed and $r_+$ is varied, whereas in the isothermal process, \eq{temperature} is inverted to obtain $P=P(r_+,T)$. The equal-$G$ construction is equivalent to the Maxwell equal-area laws
\begin{eqnarray}
&&T_*\big(S_l-S_s\big)=\int_{S_s}^{S_l}T(S,P)\mathrm{d}S,\nonumber\\
&&P_*\big(V_l-V_s\big)=\int_{V_s}^{V_l}P(V,T)\mathrm{d}V,\label{maxwell}
\end{eqnarray}
where the first and second lines apply to the isobaric $T-S$ and isothermal $P-V$ processes, respectively. The pair of temperature extrema and the equal-$G$ crossing determine the isobaric transition temperature. Figure~\ref{figthermopaths}(a) shows four isobars with $P/P_c=0.4,0.6,1.0,1.2$. The two subcritical curves oscillate, the critical curve has a horizontal inflection at $r_+=r_c$, and the supercritical curve is monotonic. Only the subcritical curves contain two stable branches separated by an unstable branch.

In the isothermal process, we fix $T$ and solve \eq{temperature} for $P$ at each radius, which gives the implicit equation of state $P=P(r_+,T)$. The implicit-function relations above show that \eq{critical} remains the correct criticality test. Figure~\ref{figthermopaths}(b) contains four isotherms with $T/T_c=0.8,0.9,1.0,1.2$. The curves below $T_c$ oscillate, the critical curve passes through the inflection point, and the supercritical curve is monotonic. The same branch counting is obtained although $P$ changes along the path.

\begin{figure}[htb]
\centering
\makebox[\linewidth][c]{%
\subfigure[a]{
\includegraphics[width=2.8in]{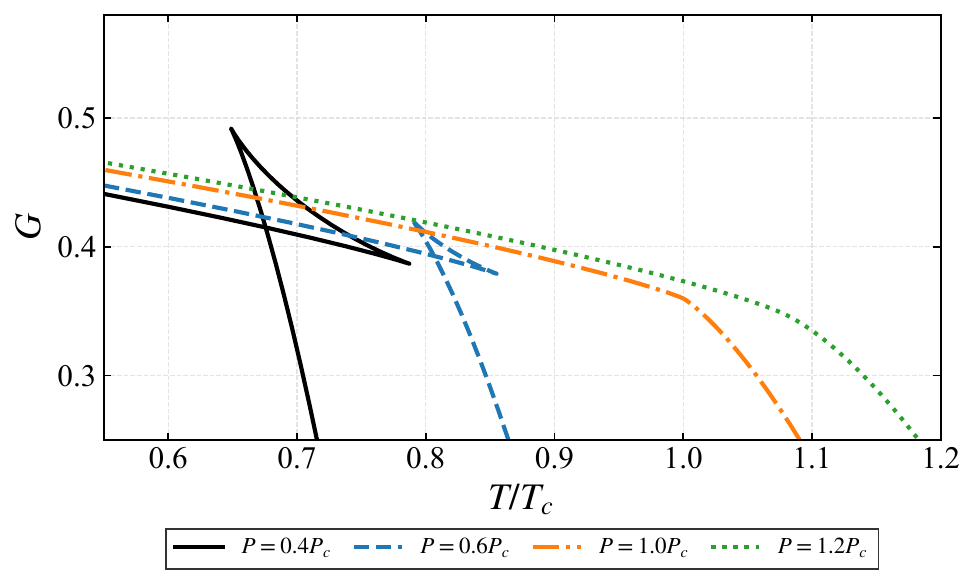}}
\hspace{0.10in}%
\subfigure[b]{
\includegraphics[width=2.8in]{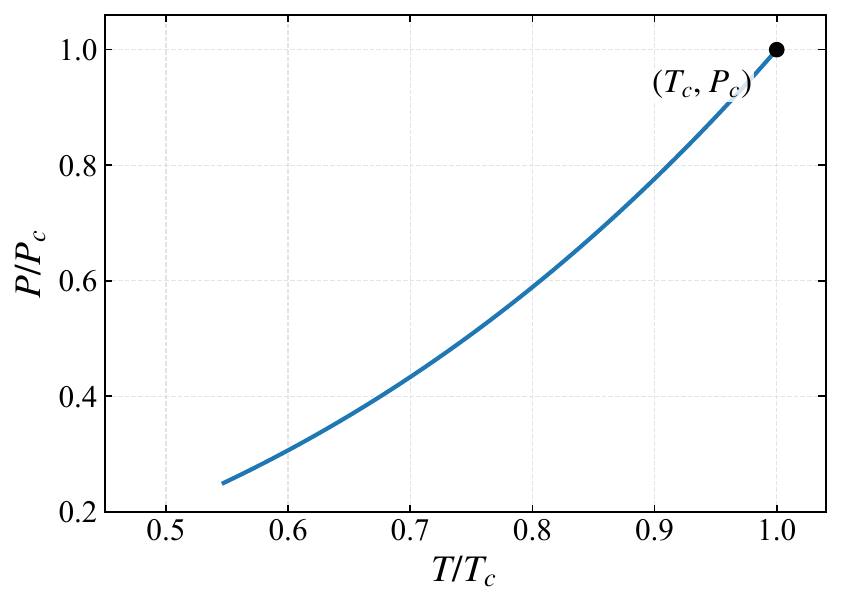}}}%
\caption{The Gibbs free energy (a) and coexistence curve (b), with fixed $\epsilon=-0.5$ and $Q_{\mathrm{m}}=0.4$. The curves in panel (a) correspond to $P/P_c=0.4,0.6,1.0,1.2$, while panel (b) displays the coexistence curve and the black point at $(T/T_c,P/P_c)=(1,1)$ marking $(T_c,P_c)$; the label is placed adjacent to the point. The swallowtail and coexistence curve terminate at the critical point.}\label{figthermoglobal}
\end{figure}

The coexistence line follows from equal temperature and equal Gibbs free energy on the small and large branches. Figure~\ref{figthermoglobal} plots this line together with the Gibbs curves. At $P=0.6P_c$, \eq{coexistence} gives $T_*/T_c=0.8065$. At fixed $T=0.8T_c$, the equal-$G$ construction gives
\begin{eqnarray}
&&P_*=0.00815518,\qquad r_s=0.946933,\nonumber\\
&&r_l=2.614718,\qquad G_*=0.393940.\label{isocoexistence}
\end{eqnarray}
Both paths give the same coexistence curve. The values $0.8065$ and $0.8$ differ because the two paths meet this curve at different points.

The multi-pressure Gibbs curves in Fig.~\ref{figthermoglobal}(a) provide a global check. We use $P/P_c=0.4,0.6,1.0,1.2$ to show two subcritical, one critical and one supercritical curve. A swallowtail appears below $P_c$ and disappears at and above $P_c$, and its crossing gives the equal-$T$, equal-$G$ condition. No crossing occurs for $\epsilon=0$ or positive coupling, confirming that one temperature minimum is insufficient for a Van der Waals transition.

\clearpage
\section{Perturbations in the Euler--Heisenberg-dilaton AdS black hole spacetime}
\label{qnm}

\subsection{Scalar field perturbation}

We now consider a massless scalar-field perturbation in the Euler--Heisenberg-dilaton AdS black-hole background \eq{metric}. In the test-field approximation, the perturbation does not modify the background geometry, and its equation of motion is the Klein--Gordon equation
\begin{eqnarray}
&&\Box\Phi=\left(-g\right)^{-1/2}\partial_\mu\left[\left(-g\right)^{1/2}g^{\mu\nu}\partial_\nu\Phi\right]=0,\label{scalarfield}\\
&&\Phi(t,r,\theta,\varphi)=\mathrm{e}^{-\mathrm{i}\omega t}Y_{lm}(\theta,\varphi)\frac{\psi(r)}{r},\label{scalaransatz}
\end{eqnarray}
where $Y_{lm}$ are spherical harmonics, $l$ is the angular quantum number and $\omega$ denotes the complex QNM frequency. 

Substituting this decomposition into the Klein--Gordon equation and using the metric \eq{metric}, we first obtain the radial equation in the coordinate $r$,
\begin{eqnarray}
&&\psi''(r)+\frac{1}{2}\Bigg[\frac{A'(r)}{A(r)}+\frac{B'(r)}{B(r)}\Bigg]\psi'(r)
 +\frac{\omega^2-V(r)}{A(r)B(r)}\psi(r)=0,\label{scalarr}
\end{eqnarray}
where a prime denotes differentiation with respect to $r$. Introducing the tortoise coordinate by
\begin{eqnarray}
&&\ud r_*=[A(r)B(r)]^{-1/2}\ud r,\label{rstar}
\end{eqnarray}
we can rewrite \eq{scalarr} in the Schr\"odinger form
\begin{eqnarray}
&&\frac{\mathrm{d}^2\psi}{\mathrm{d}r_*^2}+\Big[\omega^2-V(r)\Big]\psi=0,\label{radial}\\
&&V(r)=\frac{A'(r)B(r)+A(r)B'(r)}{2r}+\frac{A(r)}{r^2}l(l+1).\label{potential}
\end{eqnarray}
The branch trajectories and potential comparisons below use the fundamental monopole mode $l=0$, for which the centrifugal term in $V(r)$ vanishes. The convergence-controlled scan covers $l=0,1,2$ and $n=0,1,2$. Near the event horizon, $A(r)$ and $B(r)$ have simple zeros, and the surface gravity and tortoise behavior are
\begin{eqnarray}
&&\kappa_+=\frac{1}{2}\sqrt{A'(r_+)B'(r_+)},\qquad r_*\simeq \frac{1}{2\kappa_+}\ln(r-r_+),\label{surfacegravity}
\end{eqnarray}
and the physical solution is ingoing. At the AdS boundary, normalizability selects the $r^{-2}$ falloff of $\psi$. We impose
\begin{eqnarray}
&&\psi\sim \mathrm{e}^{-\mathrm{i}\omega r_*}\sim(r-r_+)^{-\mathrm{i}\omega/(2\kappa_+)}
\quad (r\rightarrow r_+),\nonumber\\
&&\psi\sim r^{-2}\quad (r\rightarrow\infty),
\end{eqnarray}
and retain only roots with $\mathrm{Im}(\omega)<0$. The real part of $\omega$ gives the oscillation frequency, whereas $|\mathrm{Im}(\omega)|$ gives the damping rate and the scalar relaxation time.

To solve \eq{radial} while imposing both boundary conditions simultaneously, we use the improved asymptotic iteration method (AIM)\cite{Cho:2009cj,Cho:2011sf}. We compactify the radial coordinate by setting
\begin{eqnarray}
&&u=1-\frac{r_+}{r},\qquad r=\frac{r_+}{1-u},\qquad 0\leq u\leq1,\nonumber\\
&&\bar A(u)=A\left[\frac{r_+}{1-u}\right],\quad
\bar B(u)=B\left[\frac{r_+}{1-u}\right],\quad
\bar V(u)=V\left[\frac{r_+}{1-u}\right].\label{compactdefs}
\end{eqnarray}
The event horizon and the AdS boundary correspond to $u=0$ and $u=1$, respectively, and a dot denotes differentiation with respect to $u$. Equation \eq{scalarr} then becomes
\begin{eqnarray}
&&\frac{\mathrm{d}^2\psi}{\mathrm{d}u^2}+p(u)\frac{\mathrm{d}\psi}{\mathrm{d}u}+q(u,\omega)\psi=0,\nonumber\\
&&p(u)=\frac{2}{u-1}+\frac{1}{2}\Bigg[\frac{\dot{\bar A}(u)}{\bar A(u)}+\frac{\dot{\bar B}(u)}{\bar B(u)}\Bigg],\nonumber\\
&&q(u,\omega)=\frac{r_+^2\big[\omega^2-\bar V(u)\big]}{(1-u)^4\bar A(u)\bar B(u)}.\label{compactradial}
\end{eqnarray}
Here $p(u)$ and $q(u,\omega)$ are the transformed coefficient functions. We remove the horizon and boundary factors by writing
\begin{eqnarray}
&&\psi(u)=u^\rho(1-u)^2\chi(u),\qquad \rho=-\frac{\mathrm{i}r_+\omega}{\sigma},\nonumber\\
&&\sigma=\sqrt{\dot{\bar A}(0)\dot{\bar B}(0)}=r_+\sqrt{A'(r_+)B'(r_+)},\label{ansatz}
\end{eqnarray}
where $\chi(u)$ is regular at both endpoints. 

Substitution into \eq{compactradial} yields
\begin{eqnarray}
&&\chi''=\lambda_0(u)\chi'+s_0(u)\chi,\label{aimstandard}\\
&&\lambda_0(u)=-\frac{1}{2}\Bigg[\frac{12}{u-1}+\frac{\dot{\bar A}}{\bar A}+\frac{\dot{\bar B}}{\bar B}\Bigg]+\frac{2\mathrm{i}r_+\omega}{\sigma u},\label{lambda0}\\
&&s_0(u)=s_{00}(u)+\omega s_{01}(u)+\omega^2s_{02}(u),\nonumber\\
&&s_{00}(u)=-\frac{1}{2}\Bigg[\frac{3\big(4\bar A+(u-1)\dot{\bar A}\big)}{(u-1)^2\bar A}-\frac{2l(l+1)}{(u-1)^2\bar B}+\frac{3\dot{\bar B}}{(u-1)\bar B}\Bigg],\nonumber\\
&&s_{01}(u)=\frac{\mathrm{i}r_+}{2\sigma}\Bigg[\frac{2(1+5u)\bar A+(u-1)u\dot{\bar A}}{(u-1)u^2\bar A}+\frac{\dot{\bar B}}{u\bar B}\Bigg],\nonumber\\
&&s_{02}(u)=\frac{r_+^2}{\sigma^2u^2}-\frac{r_+^2}{(u-1)^4\bar A\bar B}.\label{s0}
\end{eqnarray}
where all barred functions and their dots are evaluated at $u$. For the fundamental branch plots we use $l=0$; the mode scan changes $l$ explicitly in the centrifugal term and keeps the same boundary factor. We expand the coefficient functions around an interior point $u_0$, which is chosen at the maximum of the effective potential when such a maximum exists. For the large-hole branch, where the monopole potential is monotonic outside the horizon, we use the fixed value $u_0=0.32$.

Starting from $\lambda_0$ and $s_0$, the improved AIM iteration functions obey
\begin{eqnarray}
&&\lambda_n=\dot{\lambda}_{n-1}+s_{n-1}+\lambda_0\lambda_{n-1},\qquad s_n=\dot{s}_{n-1}+s_0\lambda_{n-1},\label{aimrec}
\end{eqnarray}
where dots in \eq{aimrec} denote derivatives with respect to $u$. In the improved implementation, we write
\begin{eqnarray}
&&\lambda_n(u)=\sum_{j=0}^{\infty}c_n^j(u-u_0)^j,\qquad s_n(u)=\sum_{j=0}^{\infty}d_n^j(u-u_0)^j,\nonumber\\
&&c_n^j=(j+1)c_{n-1}^{j+1}+d_{n-1}^j+\sum_{k=0}^{j}c_0^kc_{n-1}^{j-k},\nonumber\\
&&d_n^j=(j+1)d_{n-1}^{j+1}+\sum_{k=0}^{j}d_0^kc_{n-1}^{j-k}.\label{aimcoeff}
\end{eqnarray}
At the expansion point $u_0$, the QNM frequency is selected by the AIM termination condition
\begin{eqnarray}
&&\delta_n(u_0,\omega)\equiv s_n(u_0)\lambda_{n-1}(u_0)-s_{n-1}(u_0)\lambda_n(u_0)
=d_n^0c_{n-1}^0-d_{n-1}^0c_n^0=0.\label{quantization}
\end{eqnarray}
For the original branch trajectories, we calculate the termination polynomial with $n_{\max}=40$ and compare it with $n_{\max}=34$. For the coupling, angular and overtone data, we use $n_{\max}=64$ and compare it with $n_{\max}=58$. We retain roots with $\mathrm{Re}(\omega)>0$, $\mathrm{Im}(\omega)<0$ and a relative shift below $10^{-3}$. The candidate roots are obtained by the Aberth--Ehrlich method with Newton-polygon initialization.

\subsection{QNM behavior in the isobaric process}

\begin{figure}[htb]
\centering
\includegraphics[width=5.2in]{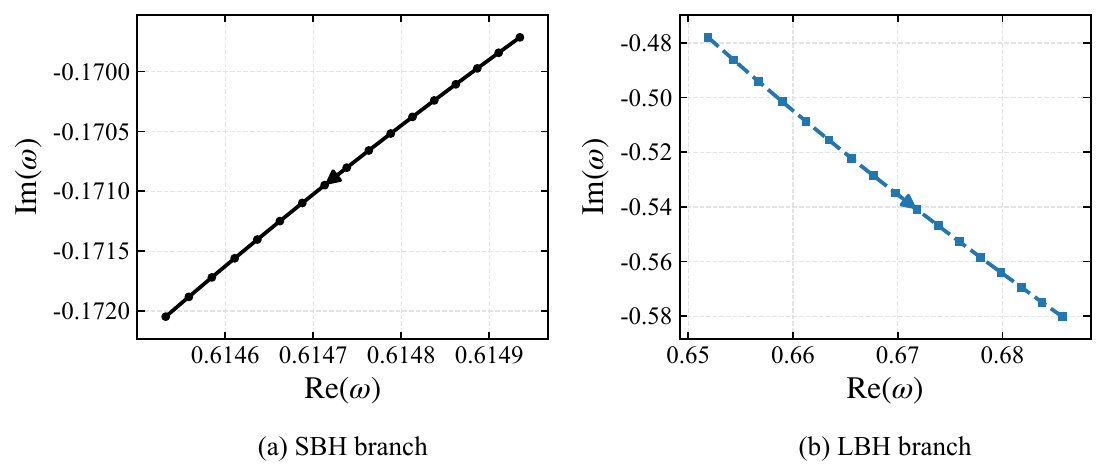}
\caption{The fundamental massless-scalar QNM trajectories in the complex-$\omega$ plane along the first-order transition isobar, with fixed $\epsilon=-0.5$, $Q_{\mathrm{m}}=0.4$, $l=0$ and $P=0.6P_c=0.00830944$. The left and right panels show the SBH and LBH branches, respectively. The arrows indicate increasing $r_+$. The separated trajectories and their different slopes provide a dynamical signature of the first-order transition.}\label{figfo}
\end{figure}

We first examine the isobaric first-order transition by fixing $\epsilon=-0.5$, $Q_{\mathrm{m}}=0.4$, $l=0$ and $P=0.6P_c$. We calculate seventeen points on each stable branch, including the coexistence endpoints in \eq{coexistence}. The SBH points satisfy $T\leq T_*$ and the LBH points satisfy $T\geq T_*$, so the metastable interval between $r_s$ and $r_l$ is excluded. All thirty-four retained roots have $\mathrm{Im}(\omega)<0$. The trajectory table and Fig.~\ref{figfo} use the baseline $N=40$ AIM data, for which the coexisting frequencies are
\begin{eqnarray}
&&\omega_{\rm SBH}=0.614532-0.172046\mathrm{i},\qquad
\omega_{\rm LBH}=0.651895-0.477955\mathrm{i}.\nonumber
\end{eqnarray}

The damping rates already differ by almost a factor of three at coexistence, showing that the two equilibrium geometries have different scalar-relaxation time scales even when their temperature and Gibbs free energy are equal. A refined calculation with $N=64$ and $N=58$ gives
\begin{eqnarray}
&&\omega_{\rm SBH}=0.614579-0.171974\mathrm{i},\qquad
\omega_{\rm LBH}=0.651895-0.477956\mathrm{i}.\label{refinedisobar}
\end{eqnarray}
The refined values in Eq.~\eqref{refinedisobar} are used in the diagnostics below.

The two trajectories in Fig.~\ref{figfo} are separated in the complex-frequency plane, and the table shows different responses of the stable phases. Along the SBH branch, increasing $T$ towards $T_*$ decreases $\mathrm{Re}(\omega)$ from $0.614935$ to $0.614532$ and changes $\mathrm{Im}(\omega)$ from $-0.169715$ to $-0.172046$. Along the LBH branch, both $\mathrm{Re}(\omega)$ and $|\mathrm{Im}(\omega)|$ increase from the coexistence point; on the displayed segment, $\mathrm{Im}(\omega)$ changes from $-0.477955$ to $-0.580137$. The smaller SBH horizon absorbs less of the perturbation, whereas the larger LBH horizon enhances absorption and damping. The branch jump is therefore visible in the monopole spectrum.

The SBH trajectory remains close to the real axis and changes only slightly, whereas the LBH trajectory lies deeper in the lower half-plane and varies more rapidly with the horizon radius. The former behavior follows from the weak change of the small-hole potential cavity, whereas the latter reflects the combined increase of horizon absorption and displacement of the AdS wall for a larger hole. Since all displayed modes have negative imaginary parts, no growing monopole mode is found, and the two-function geometry changes the spectrum quantitatively without removing the first-order QNM criterion.
\begin{table}[htb]
\centering
\caption{Fundamental massless-scalar QNMs for $l=0$ on the first-order isobar with fixed $\epsilon=-0.5$, $Q_{\mathrm{m}}=0.4$ and $P=0.6P_c$. The upper and lower blocks show the SBH and LBH branches, respectively. The displayed trajectory values use the baseline $N=40$ AIM calculation; the refined coexistence endpoint is reported separately in Table~\ref{tablev9diagnostic}.}\label{table_v9_isobar}
\begin{tabular}{c c c c}\hline\hline
$T$ & $P$ & $r_+$ & $\omega$ \\ \hline
0.0729629 & 0.00830944 & 0.951028 & $0.614532-0.172046\mathrm{i}$ \\
0.0726142 & 0.00830944 & 0.946193 & $0.614636-0.171403\mathrm{i}$ \\
0.0722655 & 0.00830944 & 0.941626 & $0.614738-0.170803\mathrm{i}$ \\
0.0719169 & 0.00830944 & 0.937297 & $0.614837-0.170241\mathrm{i}$ \\
0.0715682 & 0.00830944 & 0.933180 & $0.614935-0.169715\mathrm{i}$ \\
\hline
0.0729629 & 0.00830944 & 2.575539 & $0.651895-0.477955\mathrm{i}$ \\
0.0739514 & 0.00830944 & 2.737739 & $0.661238-0.508664\mathrm{i}$ \\
0.0749398 & 0.00830944 & 2.876518 & $0.669784-0.534858\mathrm{i}$ \\
0.0759283 & 0.00830944 & 3.001517 & $0.677889-0.558395\mathrm{i}$ \\
0.0769168 & 0.00830944 & 3.117207 & $0.685711-0.580137\mathrm{i}$ \\
\hline\hline
\end{tabular}
\end{table}

\subsection{QNM behavior in the isothermal process}

\begin{figure}[H]
\centering
\includegraphics[width=5.2in]{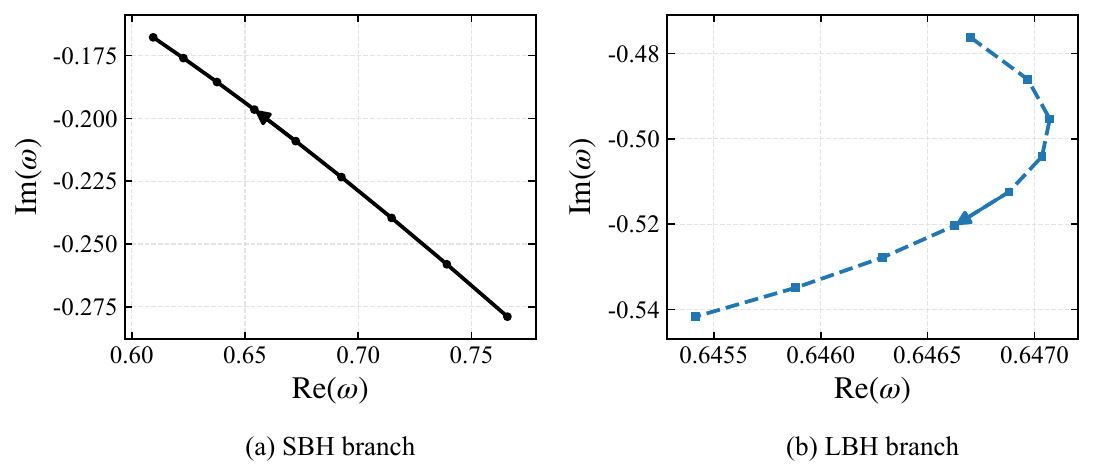}
\caption{The fundamental monopole QNM trajectories in the complex-$\omega$ plane along the first-order isotherm $T=0.8T_c$, with fixed $\epsilon=-0.5$, $Q_{\mathrm{m}}=0.4$ and $l=0$. The left and right panels show the SBH and LBH branches, respectively. The arrows indicate increasing $r_+$. The two trajectories remain discontinuous at coexistence.}\label{figisothermalqnm}
\end{figure}

We next follow the first-order transition along the isotherm $T=0.8T_c$, where the pressure is determined from the implicit equation of state. The coexistence point from \eq{isocoexistence} is $P_*=0.00815518$. We calculate nine points on each stable branch, beginning at $r_s=0.946933$ and $r_l=2.614718$. At the coexistence point, the fundamental frequencies are
\begin{eqnarray}
&&\omega_{\rm SBH}=0.609432-0.167759\mathrm{i},\qquad \omega_{\rm LBH}=0.646700-0.476223\mathrm{i}.\label{isothermalqnm}
\end{eqnarray}
The pressure increases from $P_*$ to $0.0132886$ on the SBH branch as $r_+$ decreases, whereas it decreases from $P_*$ to $0.0074028$ on the LBH branch as $r_+$ increases. The frequencies therefore trace two separated trajectories in Fig.~\ref{figisothermalqnm}, and the branch jump remains visible at fixed temperature. The opposite pressure directions change the AdS length differently on the two branches, while the radius changes the horizon absorption. Their competition on the SBH branch and partial reinforcement on the LBH branch preserve the dynamical distinction.

\begin{table}[htb]
\centering
\caption{Fundamental massless-scalar QNMs for $l=0$ on the first-order isotherm with fixed $\epsilon=-0.5$, $Q_{\mathrm{m}}=0.4$ and $T=0.8T_c$. The upper and lower blocks show the SBH and LBH branches, respectively. Five representative rows from each nine-point branch scan are displayed.}\label{table_v9_isotherm}
\begin{tabular}{c c c c}\hline\hline
$T$ & $P$ & $r_+$ & $\omega$ \\ \hline
0.0723771 & 0.00815518 & 0.946933 & $0.609432-0.167759\mathrm{i}$ \\
0.0723771 & 0.00899601 & 0.927994 & $0.637614-0.185552\mathrm{i}$ \\
0.0723771 & 0.01008658 & 0.909056 & $0.672372-0.209065\mathrm{i}$ \\
0.0723771 & 0.01149056 & 0.890117 & $0.714727-0.239652\mathrm{i}$ \\
0.0723771 & 0.01328862 & 0.871178 & $0.765859-0.278943\mathrm{i}$ \\
\hline
0.0723771 & 0.00815518 & 2.614718 & $0.646700-0.476223\mathrm{i}$ \\
0.0723771 & 0.00797658 & 2.778138 & $0.647072-0.495340\mathrm{i}$ \\
0.0723771 & 0.00778823 & 2.941558 & $0.646881-0.512426\mathrm{i}$ \\
0.0723771 & 0.00759570 & 3.104978 & $0.646290-0.527781\mathrm{i}$ \\
0.0723771 & 0.00740280 & 3.268398 & $0.645415-0.541649\mathrm{i}$ \\
\hline\hline
\end{tabular}
\end{table}

Along an isotherm, the changes of $r_+$ and $P$ are linked by the equation of state, so Fig.~\ref{figisothermalqnm} contains both geometric effects. We therefore examine the radius and pressure responses separately; the corresponding frequencies are collected in Tables~\ref{table_v9_radius_combined} and \ref{table_v9_pressure_combined}. At fixed pressure, the SBH real part decreases weakly, whereas the LBH real part increases; the modulus of the imaginary part increases on both branches. Increasing $r_+$ moves the finite AdS potential wall towards smaller $r_*$ and strengthens absorption at the horizon. The oscillation frequency is consequently enhanced on the LBH branch, and the perturbation decays faster on both branches.

\begin{figure}[H]
\centering
\includegraphics[width=5.65in]{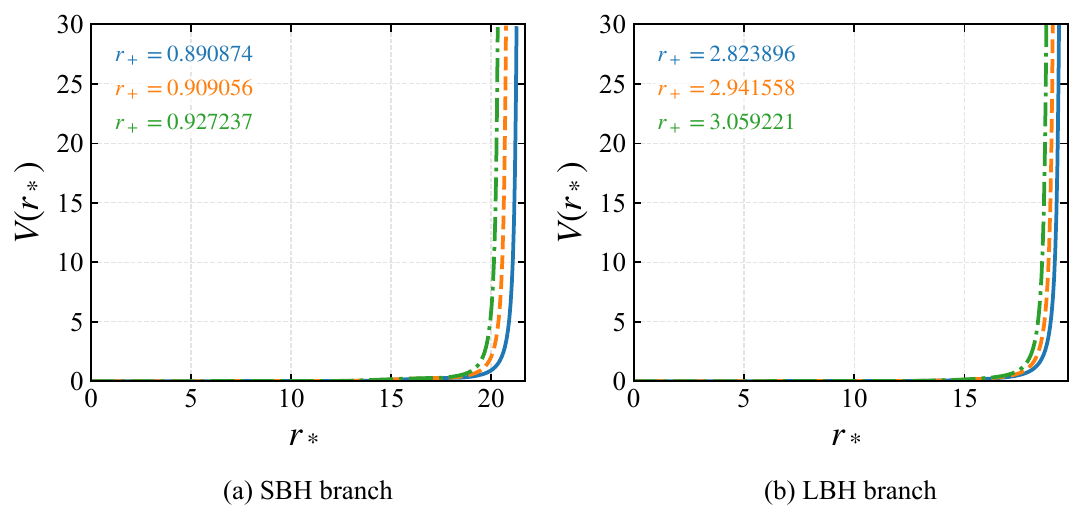}
\caption{The effective potential $V(r_*)$ for the fundamental massless-scalar perturbation ($l=0$) as a function of the tortoise coordinate $r_*$, with fixed $\epsilon=-0.5$ and $Q_{\mathrm{m}}=0.4$. The left and right panels show the SBH and LBH branches, respectively, at $P=0.01008658$ and $P=0.00778823$. The curves are labelled by $r_+=0.890874,0.909056,0.927237$ in panel (a) and $r_+=2.823896,2.941558,3.059221$ in panel (b); solid, dashed and dash-dotted lines follow the increasing-radius order. For visual clarity, each curve is displayed only up to $V=30$; the tortoise integral is evaluated with $u\in[10^{-5},0.995]$ and 9000 sample points. Increasing $r_+$ shortens the effective cavity and raises the oscillation scale.}\label{figpotentialradius}
\end{figure}

The curves in Fig.~\ref{figpotentialradius} follow the radius order stated in the caption. They stay close to zero over most of the domain and rise sharply near the AdS boundary, where $r_*$ approaches a constant. As $r_+$ increases, the wall moves towards the horizon and the effective cavity becomes shorter. This raises the oscillation scale, especially on the LBH branch. The change of the near-horizon slope also gives the larger damping modulus in the LBH block of Table~\ref{table_v9_radius_combined}.

\begin{figure}[H]
\centering
\includegraphics[width=5.65in]{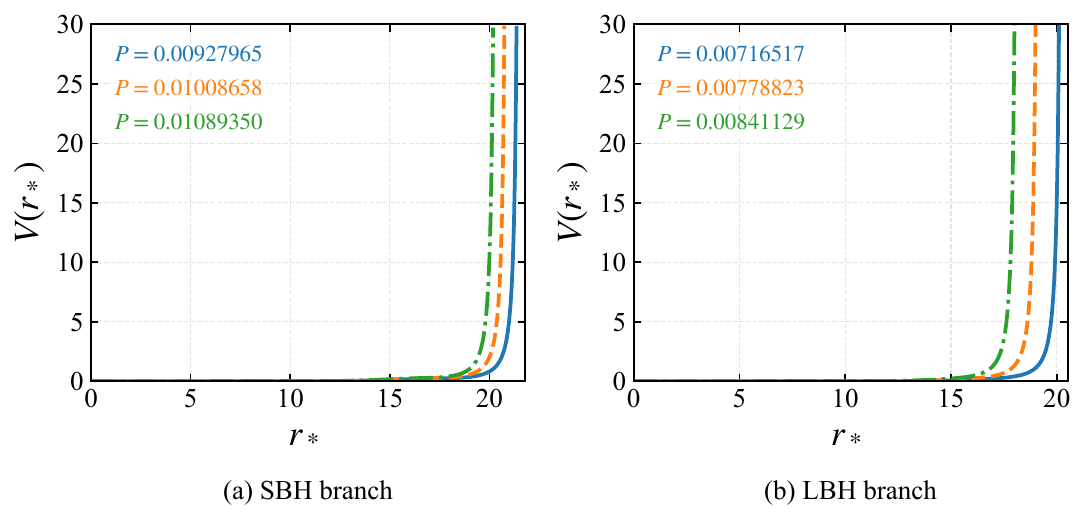}
\caption{The effective potential $V(r_*)$ for the fundamental massless-scalar perturbation ($l=0$) as a function of the tortoise coordinate $r_*$, with fixed $\epsilon=-0.5$ and $Q_{\mathrm{m}}=0.4$. The left and right panels show the SBH and LBH branches, respectively, at $r_+=0.909056$ and $r_+=2.941558$. The curves correspond to the listed pressures, with solid, dashed and dash-dotted lines in increasing-pressure order. For visual clarity, each curve is displayed only up to $V=30$; the tortoise integral is evaluated with $u\in[10^{-5},0.995]$ and 9000 sample points. Increasing $P$ moves the wall inward and raises both frequency components.}\label{figpotentialpressure}
\end{figure}

At fixed horizon radius, the pressure scans keep the horizon absorption area unchanged. The resulting QNM values are given in the two branch blocks of Table~\ref{table_v9_pressure_combined}, and the corresponding potential profiles are shown in Fig.~\ref{figpotentialpressure}.

In Fig.~\ref{figpotentialpressure}, increasing $P$ shortens the AdS length $L=(3/8\pi P)^{1/2}$ and moves the potential wall towards smaller $r_*$. Both $\mathrm{Re}(\omega)$ and $|\mathrm{Im}(\omega)|$ therefore increase in the fixed-radius tables. Along the isothermal SBH branch, this pressure effect competes with the change of $r_+$.

For a quantitative potential comparison, we record the surface gravity, the position and height of a potential maximum when it exists, and the finite AdS-wall location. For the two-function metric, these quantities are defined by
\begin{eqnarray}
&&\kappa_+=\frac{1}{2}\sqrt{A'(r_+)B'(r_+)},\qquad
V_{\rm pk}=V(r_{\rm pk}),\qquad V'(r_{\rm pk})=0,\label{potentialdiagnostic}\\
&&L_*^{(\delta,R)}=\int_{r_++\delta}^{R}\frac{\mathrm{d}r}{\sqrt{A(r)B(r)}},\qquad
V_{\rm w}(u_{\rm w})=V\!\left(\frac{r_+}{1-u_{\rm w}}\right),\quad u_{\rm w}=0.97,0.98,0.99 .\nonumber
\end{eqnarray}
The first quantity measures the absorption scale at the inner boundary, whereas $r_{\rm pk}$ and $V_{\rm pk}$ describe a finite potential barrier when a peak exists. For monotonic monopole potentials, we instead use the wall values $V_{\rm w}(u_{\rm w})$ and the tortoise length $L_*^{(\delta,R)}$. Thus, $\kappa_+$ mainly measures horizon absorption, while $L_*$ and $V_{\rm w}$ describe the AdS boundary. The cutoffs and sampling are stated in the captions.

To separate these two effects locally, we expand the frequency about a reference state,
\begin{eqnarray}
&&\omega(r_++\Delta r_+,P+\Delta P)=\omega(r_+,P)+\left(\frac{\partial\omega}{\partial r_+}\right)_P\Delta r_+\nonumber\\
&&\hspace{8mm}+\left(\frac{\partial\omega}{\partial P}\right)_{r_+}\Delta P+O(\Delta r_+^2,\Delta P^2,\Delta r_+\Delta P),\label{qnmseries}
\end{eqnarray}
where the first and second linear terms are denoted by $\Delta_r$ and $\Delta_P$, respectively. The tables quantify the independent radius and pressure responses, while the potential plots show their geometric origin. Their combined effect gives the two QNM trajectories in Fig.~\ref{figisothermalqnm}.
\begin{table}[htb]
\centering
\small
\caption{Fundamental massless-scalar QNMs for $l=0$ at fixed pressure, with $\epsilon=-0.5$ and $Q_{\mathrm{m}}=0.4$. The upper and lower blocks correspond to the SBH and LBH branches, respectively; the SBH and LBH pressures are $P=0.01008658$ and $P=0.00778823$. Five representative rows are shown for each branch.}\label{table_v9_radius_combined}
\begin{tabular}{c c c}\hline\hline
$P$ & $r_+$ & $\omega$ \\ \hline
\multicolumn{3}{c}{SBH branch}\\ \hline
0.01008658 & 0.890874 & $0.672922-0.206575\mathrm{i}$ \\ 
0.01008658 & 0.899965 & $0.672635-0.207802\mathrm{i}$ \\ 
0.01008658 & 0.909056 & $0.672372-0.209065\mathrm{i}$ \\ 
0.01008658 & 0.918146 & $0.672133-0.210366\mathrm{i}$ \\ 
0.01008658 & 0.927237 & $0.671916-0.211705\mathrm{i}$ \\ 
\hline
\multicolumn{3}{c}{LBH branch}\\ \hline
0.00778823 & 2.823896 & $0.640151-0.491591\mathrm{i}$ \\ 
0.00778823 & 2.882727 & $0.643476-0.502014\mathrm{i}$ \\ 
0.00778823 & 2.941558 & $0.646881-0.512426\mathrm{i}$ \\ 
0.00778823 & 3.000389 & $0.650366-0.522826\mathrm{i}$ \\ 
0.00778823 & 3.059221 & $0.653926-0.533216\mathrm{i}$ \\ 
\hline\hline
\end{tabular}
\end{table}
\begin{table}[htb]
\centering
\small
\caption{Fundamental massless-scalar QNMs for $l=0$ at fixed horizon radius, with $\epsilon=-0.5$ and $Q_{\mathrm{m}}=0.4$. The upper and lower blocks correspond to the SBH ($r_+=0.909056$) and LBH ($r_+=2.941558$) branches, respectively. Five representative rows are shown for each branch.}\label{table_v9_pressure_combined}
\begin{tabular}{c c c}\hline\hline
$r_+$ & $P$ & $\omega$ \\ \hline
\multicolumn{3}{c}{SBH branch}\\ \hline
0.909056 & 0.00927965 & $0.647189-0.189833\mathrm{i}$ \\ 
0.909056 & 0.00968311 & $0.659898-0.199444\mathrm{i}$ \\ 
0.909056 & 0.01008658 & $0.672372-0.209065\mathrm{i}$ \\ 
0.909056 & 0.01049004 & $0.684627-0.218694\mathrm{i}$ \\ 
0.909056 & 0.01089350 & $0.696674-0.228330\mathrm{i}$ \\ 
\hline
\multicolumn{3}{c}{LBH branch}\\ \hline
2.941558 & 0.00716517 & $0.614121-0.470919\mathrm{i}$ \\ 
2.941558 & 0.00747670 & $0.630559-0.491679\mathrm{i}$ \\ 
2.941558 & 0.00778823 & $0.646881-0.512426\mathrm{i}$ \\ 
2.941558 & 0.00809976 & $0.663096-0.533161\mathrm{i}$ \\ 
2.941558 & 0.00841129 & $0.679213-0.553884\mathrm{i}$ \\ 
\hline\hline
\end{tabular}
\end{table}

\subsection{QNM trajectory diagnostics}

\begin{figure}[htb]
\centering
\includegraphics[width=5.65in]{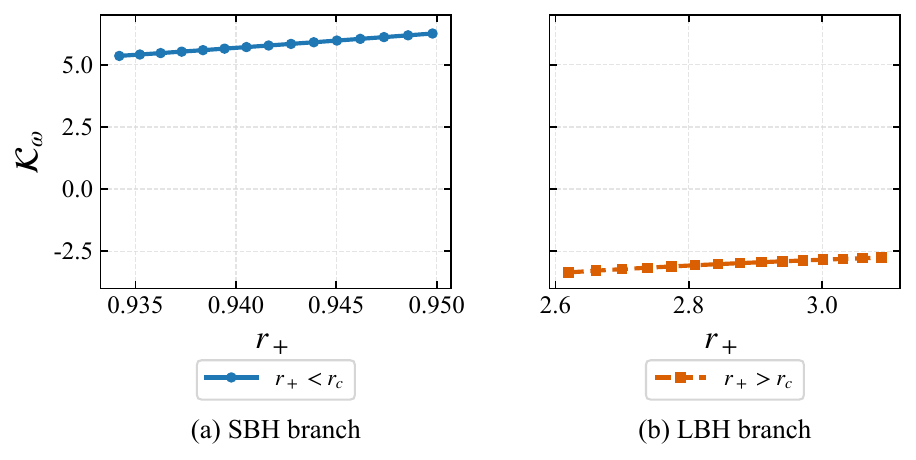}
\caption{The local QNM trajectory slope ${\cal K}_{\omega}$ as a function of $r_+$ for the first-order isobar with fixed $\epsilon=-0.5$, $Q_{\mathrm{m}}=0.4$ and $P=0.6P_c=0.00830944$. The two panels show the SBH and LBH branches, respectively; solid blue and dashed orange curves are used, and the markers show the interior AIM data points used in the centered finite-difference estimate. The trajectory data are the baseline $N=40$ values. The opposite slope signs on the two branches signal the first-order transition.}
\label{figqnmv9slope}
\end{figure}

We characterize the QNM trajectories by the local slope and the inter-branch distance:
\begin{eqnarray}
&& {\cal K}_{\omega}=\frac{\mathrm{d}\omega_{\mathrm{I}}/\mathrm{d}r_+}{\mathrm{d}\omega_{\mathrm{R}}/\mathrm{d}r_+},\qquad
D_{\omega}(T)=\left|\omega_{\mathrm{LBH}}(T)-\omega_{\mathrm{SBH}}(T)\right|.\label{qnmdiagnostic}
\end{eqnarray}
Here ${\cal K}_{\omega}$ is the slope in the complex-$\omega$ plane, and $D_{\omega}$ measures the separation of two states at the same temperature. The symbol ${\cal K}_{\omega}$ distinguishes this diagnostic from the thermodynamic function $K(r_+)$ in Sect.~\ref{thermo}. Interior derivatives use a centered finite difference on the converged AIM data, while endpoint values are omitted from Fig.~\ref{figqnmv9slope}. A small $\mathrm{d}\omega_{\mathrm{R}}/\mathrm{d}r_+$ below the frequency error is not treated as a physical divergence.

At the coexistence point on the isobar $P=0.6P_c$, the refined frequencies and their separations are
\begin{eqnarray}
&&\omega_{\rm SBH}=0.614579-0.171974\mathrm{i},\qquad
\omega_{\rm LBH}=0.651895-0.477956\mathrm{i},\nonumber\\
&&\Delta\omega_{\mathrm{R}}=0.037317,\qquad |\Delta\omega_{\mathrm{I}}|=0.305981,\qquad
D_{\omega}(T_*)=0.308248.\label{coexdiagnosticvalues}
\end{eqnarray}
The finite-difference values of ${\cal K}_{\omega}$ are positive on the SBH branch and negative on the LBH branch, with approximate ranges $5.33\leq {\cal K}_{\omega}\leq6.29$ and $-3.39\leq {\cal K}_{\omega}\leq-2.74$, respectively. The sign is reliable, whereas the endpoint values depend on the finite-difference stencil.

\begin{table}[H]
\centering
\small
\caption{Quantitative QNM diagnostics at the first-order coexistence point for $\epsilon=-0.5$, $Q_{\mathrm{m}}=0.4$, $l=0$ and $P=0.6P_c$. The SBH and LBH entries are evaluated at the common temperature $T=T_*$ using the refined $N=64$ AIM calculation, checked against $N=58$; $\Delta\omega_j$ denotes the difference between the LBH and SBH components, and $D_{\omega}$ is their complex-frequency distance.}\label{tablev9diagnostic}
\begin{tabular}{lcc}
\hline\hline
quantity & SBH & LBH \\ \hline
$r_+$ & $0.951028$ & $2.575539$ \\
$\omega_{\mathrm{R}}$ & $0.614579$ & $0.651895$ \\
$\omega_{\mathrm{I}}$ & $-0.171974$ & $-0.477956$ \\
\hline
\multicolumn{3}{c}{Derived separations} \\ \hline
$|\Delta\omega_{\mathrm{R}}|$ & \multicolumn{2}{c}{$0.037317$} \\
$|\Delta\omega_{\mathrm{I}}|$ & \multicolumn{2}{c}{$0.305981$} \\
$D_{\omega}(T_*)$ & \multicolumn{2}{c}{$0.308248$} \\
\hline\hline
\end{tabular}
\end{table}

Figure~\ref{figqnmv9slope} and Table~\ref{tablev9diagnostic} provide two complementary tests. The QNM trajectory has different local orientations on the two stable branches, whereas the finite value of $D_{\omega}$ gives a branch-independent measure of the frequency jump. At the critical point, the two branches should be replaced by one curve, so $D_{\omega}\rightarrow0$ when the two states are selected at the same critical temperature. This statement concerns the limit of the branch construction and does not imply that the derivative ${\cal K}_{\omega}$ must remain finite for every mode.

\section{QNM behavior at the critical point}
\label{critqnm}

\subsection{Critical trajectories}

At the critical point, the two temperature extrema coalesce and the SBH and LBH branches form one continuous path. We set $\epsilon=-0.5$, $Q_{\mathrm{m}}=0.4$ and $l=0$, and calculate twenty-six states on the critical isobar and eleven states on the critical isotherm. The two paths give the same frequency within $2.5\times10^{-6}$,
\begin{eqnarray}
&&\omega_c=0.790380-0.417249\mathrm{i}.\label{criticalqnm}
\end{eqnarray}
The value in Eq.~\eqref{criticalqnm} is taken from the critical-isobar data. The two paths provide a numerical check: the isobar varies $T$ at fixed $P$, whereas the isotherm varies $P$ at fixed $T$.

\begin{figure}[H]
\centering
\includegraphics[width=5.65in]{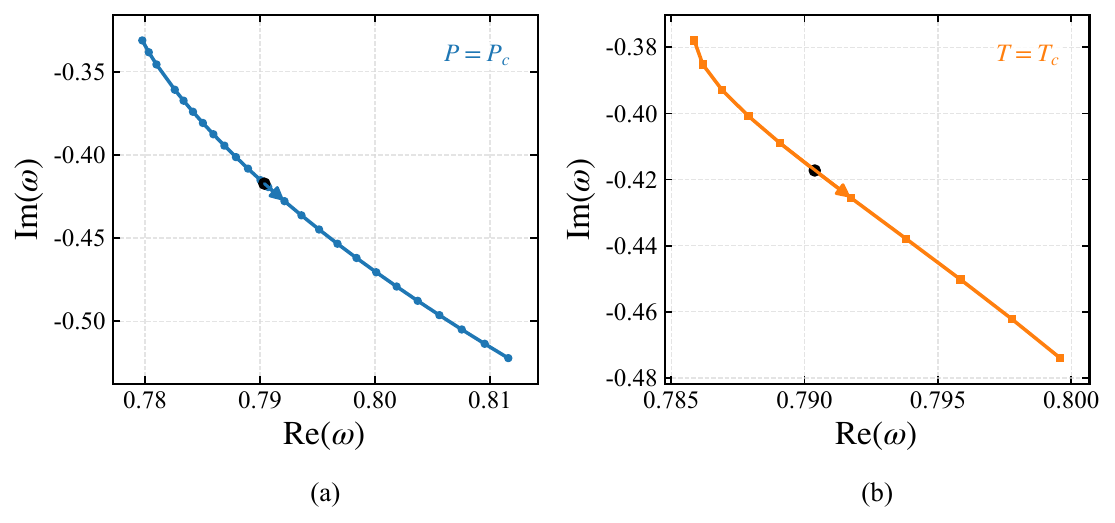}
\caption{The fundamental monopole QNM trajectories in the complex-$\omega$ plane at the critical point, with fixed $\epsilon=-0.5$, $Q_{\mathrm{m}}=0.4$, $l=0$ and $r_c=1.34693$. Panel (a) is the critical isobar $P=P_c=0.0138491$, and panel (b) is the critical isotherm $T=T_c=0.0904713$. The arrows indicate increasing $r_+$, the black points mark the critical frequency $\omega_c=0.790380-0.417249\mathrm{i}$, and both trajectories are continuous through the critical state.}\label{figcriticalcombined}
\end{figure}

For the critical isobar, the fixed pressure and sampled interval are
\begin{eqnarray}
&&P=P_c=0.0138491,\qquad r_c=1.34693,\qquad 1.05060\leq r_+\leq1.68366,\nonumber\\
&&0.0884448\leq T\leq0.0911496.\label{criticalisobar}
\end{eqnarray}
Here the first line gives the critical pressure and radius, and the second line gives the temperature range. Figure~\ref{figcriticalcombined}(a) shows one QNM trajectory through $\omega_c$. As $r_+$ increases, $\mathrm{Re}(\omega)$ grows from $0.779768$ to $0.811580$, while $\mathrm{Im}(\omega)$ changes from $-0.331127$ to $-0.522185$. Thus, the oscillation frequency and damping rate both increase along this continuous path. The change follows from the deformation of the effective potential cavity after the two thermodynamic branches have merged.

For the critical isotherm, we invert the temperature equation and obtain
\begin{eqnarray}
&&T=T_c=0.0904713,\qquad 0.0137987\leq P\leq0.0138972.\label{criticalisotherm}
\end{eqnarray}
The pressure decreases as $r_+$ increases. The trajectory in Fig.~\ref{figcriticalcombined}(b) reaches the same frequency in Eq.~\eqref{criticalqnm}. Along this path, $\mathrm{Re}(\omega)$ changes from $0.785860$ to $0.799590$ and $\mathrm{Im}(\omega)$ from $-0.377912$ to $-0.474001$. The fixed temperature is maintained by the accompanying change of the AdS scale, so both components remain continuous through $r_+/r_c=1$.

At criticality, one smooth monopole QNM curve replaces the two separated stable-branch trajectories below $P_c$, and the Gibbs swallowtail disappears at the same point. The present $l=0$ calculation shows continuity, but it does not imply a dynamical singularity for every mode.

\subsection{Diagnostics across regimes and modes}

For each subcritical pressure, the equal-$G$ condition gives a coexistence temperature. We quantify the separation of the two equilibrium responses by
\begin{eqnarray}
&&D_{\omega}^{\rm coex}(P)=\left|\omega_{\rm LBH}(T_*(P),P)-\omega_{\rm SBH}(T_*(P),P)\right|,\qquad P<P_c.\label{coexdistance}
\end{eqnarray}
Here $D_{\omega}^{\rm coex}$ is the complex-frequency distance at equal temperature. A finite value together with two stable branches indicates the first-order regime. It tends to zero at $P_c$, where the response becomes single-valued. For $P=0.6P_c$, the refined monopole data give
\begin{eqnarray}
&&D_{\omega}^{\rm coex}=0.308248,\qquad {\cal R}_{\omega}=0.213096.\label{coexcontrast}
\end{eqnarray}
The critical isobar and isotherm pass through one common frequency within the AIM uncertainty.

We separate coupling effects from coexistence by evaluating a common state with $P=0.01$, $r_+=1$ and $Q_{\mathrm{m}}=0.4$. The modes for $l=0,1,2$ are listed in Table~\ref{table_v9_epsilon_modes}. For $l=0$, $\mathrm{Re}(\omega)$ changes from $0.668674$ at $\epsilon=-1$ to $0.680335$ at $\epsilon=1$, while $|\mathrm{Im}(\omega)|$ decreases from $0.233071$ to $0.186078$. Increasing $\epsilon$ changes the short-distance term in $A(r)$ and hence the horizon absorption, giving a longer scalar-relaxation time at this fixed state. The normalized contrast is defined as
\begin{eqnarray}
&&{\cal R}_{\omega}(P,\epsilon,Q_{\mathrm{m}})=
\frac{D_{\omega}^{\rm coex}(P)}{|\omega_{\rm SBH}(T_*,P)|+|\omega_{\rm LBH}(T_*,P)|}.\label{contrastmap}
\end{eqnarray}
The common-state scan is a geometry comparison rather than a coexistence test. For negative $\epsilon$, the finite-radius critical points and their fundamental modes are listed in Table~\ref{table_v9_critical_modes}. No critical point is listed for $\epsilon=0$ or the tested positive couplings because the thermodynamic scan gives no equal-$G$ SBH/LBH crossing.
\begin{table}[H]
\centering
\small
\caption{Converged fundamental QNMs ($n=0$) for the common state $P=0.01$, $r_+=1$ and $Q_{\mathrm{m}}=0.4$, with $\epsilon$ varied from $-1$ to $1$. The three frequency columns correspond to $l=0,1,2$, respectively. The AIM order is $N=64$, and the relative shift from $N=58$ is smaller than $10^{-3}$ for every entry.}\label{table_v9_epsilon_modes}
\begin{tabular}{c c c c c}\hline\hline
$\epsilon$ & $T$ & $\omega_{l=0,n=0}$ & $\omega_{l=1,n=0}$ & $\omega_{l=2,n=0}$ \\ \hline
$-1.0$ & $0.0685718$ & $0.668674-0.233071\mathrm{i}$ & $0.918868-0.125281\mathrm{i}$ & $1.232194-0.055955\mathrm{i}$ \\
$-0.5$ & $0.0788933$ & $0.668134-0.221561\mathrm{i}$ & $0.922588-0.118048\mathrm{i}$ & $1.237287-0.051079\mathrm{i}$ \\
$-0.1$ & $0.0871573$ & $0.669492-0.211581\mathrm{i}$ & $0.926082-0.112261\mathrm{i}$ & $1.241598-0.047333\mathrm{i}$ \\
$0$ & $0.0892243$ & $0.670091-0.209083\mathrm{i}$ & $0.927030-0.110831\mathrm{i}$ & $1.242706-0.046423\mathrm{i}$ \\
$0.1$ & $0.0912918$ & $0.670786-0.206606\mathrm{i}$ & $0.928007-0.109411\mathrm{i}$ & $1.243827-0.045525\mathrm{i}$ \\
$0.5$ & $0.0995659$ & $0.674401-0.197027\mathrm{i}$ & $0.932186-0.103866\mathrm{i}$ & $1.248419-0.042064\mathrm{i}$ \\
$1.0$ & $0.1099189$ & $0.680335-0.186078\mathrm{i}$ & $0.937940-0.097339\mathrm{i}$ & $1.254364-0.038060\mathrm{i}$ \\
\hline\hline
\end{tabular}
\end{table}
\begin{table}[H]
\centering
\scriptsize
\caption{Converged fundamental QNMs ($n=0$) at the finite-radius critical points for $Q_{\mathrm{m}}=0.4$. The thermodynamic triples in the second column are $(r_c,P_c,T_c)$, and the frequency columns correspond to $l=0,1,2$, respectively. The AIM order is $N=64$, with relative shifts below $10^{-3}$ from $N=58$.}\label{table_v9_critical_modes}
\setlength{\tabcolsep}{2pt}
\begin{tabular}{c c c c c}\hline\hline
$\epsilon$ & $(r_c,P_c,T_c)$ & $\omega_{l=0,n=0}$ & $\omega_{l=1,n=0}$ & $\omega_{l=2,n=0}$ \\ \hline
$-1.0$ & $(1.489387,0.0116781,0.0834539)$ & $0.728855-0.387953\mathrm{i}$ & $0.957793-0.295248\mathrm{i}$ & $1.256777-0.215032\mathrm{i}$ \\
$-0.5$ & $(1.346927,0.0138491,0.0904713)$ & $0.790380-0.417249\mathrm{i}$ & $1.041585-0.315318\mathrm{i}$ & $1.368667-0.227852\mathrm{i}$ \\
$-0.1$ & $(1.077257,0.0197503,0.1066531)$ & $0.932596-0.480063\mathrm{i}$ & $1.239356-0.354928\mathrm{i}$ & $1.635274-0.250188\mathrm{i}$ \\
\hline\hline
\end{tabular}
\end{table}

We compare angular and overtone modes at the same $\epsilon=-0.5$ coexistence point. The fundamental frequency distances and normalized contrasts are
\begin{eqnarray}
&&D_{\omega}^{\rm coex}\big|_{l=0,1,2}=0.308248,\ 0.341660,\ 0.346669,\nonumber\\
&&{\cal R}_{l,n=0}=0.213096,\ 0.191897,\ 0.153308.\label{modecontrasts}
\end{eqnarray}
The first line corresponds to $l=0,1,2$, respectively, and the second line gives the corresponding normalized values. The angular barrier increases the absolute frequency scale but reduces the relative phase contrast. The corresponding QNMs are listed in Table~\ref{table_v9_coexistence_modes}. The convergence indicator used to select the retained roots is
\begin{eqnarray}
&&\epsilon_{\rm AIM}(l,n)=
\frac{|\omega_{N}(l,n)-\omega_{N-\Delta N}(l,n)|}{|\omega_{N}(l,n)|}<10^{-3},\qquad \Delta N=6.\label{aimerrorv9}
\end{eqnarray}
Here $N$ denotes the AIM iteration order. We retain the grid $l=0,1,2$ and $n=0,1,2$ only when Eq.~\eqref{aimerrorv9} is satisfied; 61 of 63 tested $(\epsilon,l,n)$ combinations pass this test, and the omitted roots are listed in Table~\ref{table_v9_coexistence_modes}. The unstable $n=2$ roots are excluded from the coexistence contrast, since termination-polynomial root exchange is then mixed with overtone dependence.
\begin{table}[H]
\centering
\small
\caption{Converged QNMs at the first-order coexistence point, with $\epsilon=-0.5$, $Q_{\mathrm{m}}=0.4$, $P=0.00830944=0.6P_c$ and $T=T_*$. The rows give the angular number and overtone. The SBH and LBH columns use $r_s=0.951028$ and $r_l=2.575539$, respectively, and the AIM order is $N=64$ with a relative shift below $10^{-3}$ from $N=58$. A dash means that the corresponding $n=2$ root did not pass this convergence test.}\label{table_v9_coexistence_modes}
\begin{tabular}{c c c c}\hline\hline
$l$ & $n$ & $\omega_{\mathrm{SBH}}$ & $\omega_{\mathrm{LBH}}$ \\ \hline
0 & 0 & $0.614579-0.171974\mathrm{i}$ & $0.651895-0.477956\mathrm{i}$ \\
0 & 1 & $1.030399-0.390440\mathrm{i}$ & $1.104632-0.921729\mathrm{i}$ \\
0 & 2 & {\textemdash} & $1.559416-1.362241\mathrm{i}$ \\
1 & 0 & $0.859350-0.076565\mathrm{i}$ & $0.818113-0.415728\mathrm{i}$ \\
1 & 1 & $1.207030-0.279294\mathrm{i}$ & $1.234098-0.867133\mathrm{i}$ \\
1 & 2 & $1.600836-0.494753\mathrm{i}$ & $1.669714-1.313440\mathrm{i}$ \\
2 & 0 & $1.155889-0.023704\mathrm{i}$ & $1.047254-0.352911\mathrm{i}$ \\
2 & 1 & $1.460984-0.171755\mathrm{i}$ & $1.424510-0.795854\mathrm{i}$ \\
2 & 2 & {\textemdash} & $1.835837-1.242876\mathrm{i}$ \\
\hline\hline
\end{tabular}
\end{table}

\begin{samepage}
The critical point also permits a scaling test. At $P=P_c$, the expansion is
\begin{eqnarray}
&&T-T_c\propto(r_+-r_c)^3.\label{cubicthermo}
\end{eqnarray}
\end{samepage}
An analytic AIM root without mode exchange then obeys
\begin{eqnarray}
&&|\omega_j-\omega_{c,j}|=A_j\left|\frac{T-T_c}{T_c}\right|^{1/3}
\left[1+B_j\left|\frac{T-T_c}{T_c}\right|^{1/3}+\cdots\right],\qquad j=\mathrm{R,I},\nonumber\\
&&|\omega_j-\omega_{c,j}|=\widetilde A_j\left|\frac{P-P_c}{P_c}\right|^{1/3}+\cdots
\quad (T=T_c).\label{criticalscaling}
\end{eqnarray}
Here $j=\mathrm{R,I}$ labels the real and imaginary components. The exponent $1/3$ follows from the cubic thermodynamic inflection and is not assumed as a QNM property. The fitted values are
\begin{eqnarray}
&&\nu_{|\omega|}^{(P=P_c)}=0.3284\pm0.0107,\qquad R^2=0.9762,\nonumber\\
&&\nu_{|\omega|}^{(T=T_c)}=0.3319\pm0.0220,\qquad R^2=0.9660.\label{criticalfitvalues}
\end{eqnarray}
They support cubic scaling for the present monopole over the fitted intervals, but they do not establish universality; see Table~\ref{table_v9_critical_scaling}.
\begin{table}[H]
\centering
\small
\caption{Log--log fits of the critical QNM displacement for $\epsilon=-0.5$, $Q_{\mathrm{m}}=0.4$ and $l=0$. The isobar uses the variable $|T-T_c|/T_c$ and 25 noncritical points, while the isotherm uses $|P-P_c|/P_c$ and 10 noncritical points. The quoted standard errors are obtained from the ordinary least-squares slope covariance. The last column is the squared correlation coefficient for $|\omega-\omega_c|$.}\label{table_v9_critical_scaling}
\begin{tabular}{c c c c c c}\hline\hline
path & variable & $\nu_{\mathrm{R}}$ & $\nu_{\mathrm{I}}$ & $\nu_{|\omega|}\pm\sigma_\nu$ & $R^2$ \\ \hline
$P=P_c$ & $|T-T_c|/T_c$ & $0.3257$ & $0.3284$ & $0.3284\pm0.0107$ & $0.9762$ \\
$T=T_c$ & $|P-P_c|/P_c$ & $0.3069$ & $0.3324$ & $0.3319\pm0.0220$ & $0.9660$ \\
\hline\hline
\end{tabular}
\end{table}

\section{Conclusion and discussion}
\label{con}

In this paper, we have studied the horizon structure, extended thermodynamics and scalar QNMs of magnetically charged Euler--Heisenberg-dilaton AdS black holes. The sign of $\epsilon$ fixes the small-radius behavior: for $\epsilon<0$, a finite extremal radius and a two-horizon branch are present, while the $\epsilon=0$ limit reaches the GHS threshold at $r=0$ and the positive couplings examined here give one outer horizon. By varying $P$ and $\epsilon$, we obtain the first law and the Smarr relation with the conjugate pair $(\epsilon,\Psi_{\epsilon})$.

On the physical branch $K(r_+)>Q_{\mathrm{m}}^2$, the negative-coupling case has a finite critical point and a Gibbs swallowtail below $P_c$. For $\epsilon=-0.5$ and $Q_{\mathrm{m}}=0.4$, the critical values used in the QNM analysis are
$r_c=1.34693,\qquad P_c=0.0138491,\qquad T_c=0.0904713$.
The GHS limit and the tested positive couplings have no finite-radius Van der Waals-like critical point. Below $P_c$, the improved AIM gives separated SBH and LBH trajectories with opposite local slopes, reflecting the different effective potential cavities and horizon absorption scales of the two phases. At criticality, the isobaric and isothermal paths join into one curve through the critical values. The fitted displacement exponents are close to $1/3$ over the stated windows, as expected from the cubic thermodynamic inflection.

The mode scan is restricted to test-field scalar perturbations with $Q_{\mathrm{m}}=0.4$ and the listed $l,n$ values. The angular barrier changes the absolute spectrum and reduces the normalized phase contrast as $l$ increases, so the present dynamical diagnostic should be regarded as a monopole test-field result. Electromagnetic and gravitational perturbations, a wider $(\epsilon,Q_{\mathrm{m}})$ phase diagram, and related observables such as greybody factors, particle motion and shadows remain to be studied in future work.

\hspace*{3em}

{\bf Acknowledgments:}
We appreciate Chao-Ming Zhang and Lina Zhang for helpful discussions. 
This research is supported by the Jiangxi Provincial
Natural Science Foundation (Grant No. 20262BAC240347), National Natural Science Foundation of China
(NNSFC) (Grant No. 12365009) and the Natural Science Basic Research Program of
Shaanxi Province (Program No. 2026JC-YBMS-0083).

\hspace*{3em}

{\bf Data Availability Statement:}
Data sharing is not applicable to this article as no new data were created or analyzed in this study.

\clearpage
\bibliographystyle{unsrt}
\bibliography{refs}

\end{document}